\documentclass[amsmath,nofootinbib,notitlepage,prd,superscriptaddress,twocolumn]{revtex4-2}

\usepackage[T1]{fontenc}
\usepackage[usenames,dvipsnames]{xcolor}
\usepackage{aas_macros,graphicx,hyperref,orcidlink}
\hypersetup{colorlinks=true,citecolor=RoyalBlue,linkcolor=RoyalBlue,urlcolor=RoyalBlue}

\newcommand{\br}[1]{\left[#1\right]}

\newcommand{\pa}[1]{\left(#1\right)}
\newcommand{\ed}{\mathop{}\!\mathrm{d}}
\DeclareMathOperator\arcsinh{arcsinh}

\usepackage[T1]{fontenc}
\begin{document}

\title{Inferring Black Hole Spin from Interferometric Measurements of the First Photon Ring: A Geometric Approach}

\author{Lennox S. Keeble\,\orcidlink{0009-0009-5796-631}}
\email{lkeeble@princeton.edu}
\affiliation{Department of Physics, Princeton University, Princeton, NJ 08544, USA}

\author{Alejandro C\'ardenas-Avenda\~no\,\orcidlink{0000-0001-9528-1826}}
\affiliation{Computational Physics and Methods (CCS-2), Center for Nonlinear Studies (CNLS) \& Center for Theoretical Astrophysics (CTA), Los Alamos National Laboratory, Los Alamos NM 87545, USA}

\author{Daniel C. M. Palumbo\orcidlink{0000-0002-7179-3816}}
\affiliation{Center for Astrophysics $\arrowvert$ Harvard \& Smithsonian, 60 Garden Street, Cambridge, MA 02138, USA}
\affiliation{Black Hole Initiative at Harvard University, 20 Garden Street, Cambridge, MA 02138, USA}

\begin{abstract}
Accurately inferring black hole spin is crucial for understanding black hole dynamics and their astrophysical environments. In this work, we outline a geometric method for spin estimation by using the interferometric shape of the first photon ring ($n=1$) as an approximation to the critical curve, which, given an assumed value of the black hole inclination, is then mapped to a spin value. While future space‐based missions will capture a wealth of data on the first photon ring—including the full angle‐dependent diameter, angular brightness profile, and astrometric offset from the $n=0$ ring—our analysis is restricted to using only two angle-dependent diameters to compute its shape asymmetry and infer spin. Focusing on low inclinations and moderate‐to‐high spins, we test the method across various emission models, baselines, and noise sources, including a mock space-based observation. Although the size of the $n=1$ ring depends on the emission model, its interferometric shape remains a robust spin probe at low inclinations. We find that the inferred asymmetry of the $n=1$ image may be biased by the critical curve morphology, and it can be heavily skewed by the presence of noise, whether astrophysical or instrumental. In low-noise limits at low viewing inclination, significant contributions from the $n=0$ image at short baselines may lead to a downward bias in asymmetry estimates. While our method can estimate high spins in noise-free time-averaged images, increasing the noise and astrophysical variability degrades the resulting constraints, providing only lower bounds on the spin when applied to synthetic observed data. Remarkably, even using only the ring’s asymmetry, we can establish lower bounds on the spin, underscoring the promise of photon ring‐based spin inference in future space-based very long baseline interferometry missions, such as the proposed Black Hole Explorer.
\end{abstract}

\maketitle

\section{Introduction}

Astrophysical black holes are thought to be characterized by two parameters: their mass and angular momentum. That is, from the smallest to the largest black holes in the universe, just two parameters fully describe the geometry around them. Measuring these parameters for astrophysical black holes is paramount to modern astrophysics, but is a highly non-trivial task. 

Several black hole mass measurements have been carried out to date with differing levels of precision. For certain black holes, like Sgr A*, the supermassive black hole (SMBH) at the center of our galaxy, the mass has been measured with exquisite precision~\cite{GRAVITY:2021xju}. The mass of M87*, the SMBH at the center of the galaxy Messier 87, has been measured within a $\sim 20\%$ error~\cite{Gebhardt:2011yw,EventHorizonTelescope:2019ggy}. In contrast, spin measurements have proved more difficult~\cite{EventHorizonTelescope:2022wkp,EventHorizonTelescope:2024dhe}.

Using gravitational observations, the spin of a \emph{remnant} black hole from a binary black hole merger can be fairly well-measured~\cite{KAGRA:2021duu}. In fact, for several binary configurations, including two non‐spinning black holes of comparable mass, the spin is always around $a_{*}=0.7$, due to balance between the orbital angular momentum that remains at the time of merger and the fraction of angular momentum radiated away. The individual spins of the pre-merger binary, on the other hand, are very challenging to measure. This is because during the inspiral, we are better able to measure a combination of their individual spins~\cite{Purrer:2015nkh}. 

It is more challenging to measure spin via electromagnetic observations~\cite{Brenneman:2013oba,Reynolds:2020jwt}. This is because the signal we observe depends on \emph{both} the theory of gravity and the properties of the underlying plasma which produces the light we observe. This is in sharp contrast with gravitational-wave measurements, where the systematic error comes from the methods employed in solving the Einstein field equations, and the range of physical effects one is (and is not) including when building waveform templates (e.g., whether or not one accounts for eccentricity or precession effects). On the electromagnetic side, the systematic errors stem from the uncertainty in the specific properties of the surrounding material (such as, for example, the composition and geometry of the emission region or its magnetization state). Despite these tremendous challenges posed by the uncertainty in the details of the astrophysical environment surrounding BHs, spin measurements have been provided for several astrophysical sources, including supermassive black holes, where we are currently blind in the gravitational spectrum~\cite{LISA:2022kgy}.

The seminal measurements by the Event Horizon Telescope (EHT)~\cite{EventHorizonTelescope:2019dse,EventHorizonTelescope:2022wkp,EventHorizonTelescope:2024dhe} have opened up a new window into spin measurements of supermassive black holes~\cite{Johnson:2023ynn,Lupsasca:2024xhq}. Current measurements only provide weak bounds on the spin of these EHT sources~\cite{EventHorizonTelescope:2022wkp,EventHorizonTelescope:2024dhe}. It is very likely that \emph{precision} spin measurements will require observations of high spatial frequency Fourier components not accessible on Earth-based baselines, but requiring space-based very long baseline interferometry (VLBI)~\cite{Johnson:2024ttr,Lupsasca:2024xhq}.

The interferometric signal we observe from these supermassive black holes, is sourced by photons emitted in its surrounding emission. These photons execute $n$ half orbits around the black hole before either falling in or escaping to infinity and reaching our detectors. In the image domain, $n=0$ corresponds to a direct image of the emission, while each successive $n$ corresponds to a distinct ring-like image of the emission which is exponentially thinner than the last. As $n$ increases, the shape of each photon ring becomes less astrophysics dependent and asymptotes, in the limit $n\to\infty$, to the (theoretical) critical curve: an infinitely thin curve in the observer sky whose shape depends only on spin of the black hole and the observer's inclination relative to the spin axis~\cite{Bardeen1973}. 

Some observational efforts aimed at measuring the photon ring are expected to be conducted directly in the Fourier domain. In this approach, the signal does not maintain the clear (theoretical) separation of a total image into its constituent sub-images, labeled as $n=0, 1,\ldots$. Rather, the total interferometric signal is composed of interfering signals of the distinct photon rings. In certain regions, due to the hierarchical thickness of these rings, only the signal from a single ring dominates, allowing for a measurement of the $n$th photon ring. In so-called transition regions~\cite{Paugnat2022,CardenasAvendano_2024}, the $n$th photon ring's interferometric signal is of the same order of magnitude as that of $n+1$, and their interference prevents a clean measurement of either. The structure of the interferometric signal consists of a succession of these regions of domination and transition: the $n=0$ contribution dominates on very short baselines, then there is a transition to the $n=1$ dominated region and so on. In order to measure higher-order photon rings, then, one must carry out the observations at longer and longer baselines, which is observationally challenging.

A precision measurement of spin becomes increasingly simple at higher baselines since the degree of coupling of the interferometric signal to the complex astrophysics washes out. Since the higher-order photon rings asymptote to the astrophysics-independent critical curve, such measurements would provide a cleaner probe of the geometry around the black hole~\cite{Johnson_2020}. Indeed, the shape of the critical curve (in units of the BH mass) is a function of only the spin, $a_{*}=|J|/M$, of the black hole and the inclination, $i_{*}$, of the observer relative to its spin axis. Its shape can therefore be seen as a mapping to the spin and inclination of the underlying black hole. 

In Ref.~\cite{Farah:2020jkv} it was shown how polar curves provide a low-dimensional geometric representation of the black hole critical curve, enabling parametric model fitting, diagnostics of its properties, and assessments of their consistency with the Kerr solution. While accurately knowing its shape constrains mass, spin, and inclination only to within degeneracies, independent measurements of any one parameter can resolve the others, making possible increasingly precise analyses for future observations.

The critical curve is, however, \emph{unobservable}; it is infinitely thin and so has no observable interferometric signal. Remarkably, it has been shown (e.g., Refs.~\cite{GLM2020,Paugnat2022,Gralla_2020}) that the shape of the $n=2$ photon ring approximates the shape of the critical curve very well---on the sub percent order---for a wide range of astrophysical emission profiles. Consequently, a precision measurement of the shape of the $n=2$ photon ring could be used as an \emph{approximation} of the critical curve shape, which, in turn, could itself be used to infer one or both of the underlying geometric black hole parameters $(a_{*},i_{*})$, depending on whether its mass $M$ is known.

Unfortunately, the $n=2$ ring is already a challenging target for observations. This is because, for a broad class of simulations of both M$87^{*}$ and Sgr A$^{*}$, its interferometric signal is expected to dominate at (very!) long baselines ($\gtrsim 150$ G$\lambda$). For example, for observations at $345$ GHz, a baseline of $150$ G$\lambda$ corresponds to a telescope separation of approximately $10$ Earth diameters, making such measurements very challenging (and expensive) with current technology. 

The current target for planned space-based VLBI missions is the first ($n=1$) photon ring, for which the interferometric signature dominates on earlier baselines, but which is more coupled to the details of the astrophysics surrounding the black hole than the $n=2$ ring. Nevertheless, it was shown in Ref.~\cite{CardenasAvendano_2023} that for a large number of astrophysical models of the surrounding emission, the interferometric signature of the first photon ring exhibits a characteristic ringing whose periodicity defines a shape (inferred from the interferometric diameters) in visibility space, which closely tracks that of the critical curve at low inclinations. In Ref.~\cite{CardenasAvendano_2024}, an inexpensive curve fitting scheme was used to infer the shape of the first photon ring in the presence of phenomenologically-modeled instrumental noise and astrophysical fluctuations, serving as a proof-of-concept that observations of the shape of the first-photon ring are feasible in practice, even when employing simple data analysis techniques.

The spin inference method examined in this paper is motivated by the above discussion. The method itself consists of using photon ring shapes inferred from black hole interferometric signals to approximate the shape of the critical curve, which is then mapped onto a value of spin. Both the shape inference scheme, an extension of that employed in Refs.~\cite{CardenasAvendano_2024, Paugnat2022}, and the close tracking of the $n=2$ ring shape to the critical curve rest on the assumption of low observer inclination ($i_{*}\lesssim{45}$)~\cite{Paugnat2022}.

By analytically computing critical curve shapes across the parameter space $0\leq a_{*}<1$, $0^{\circ}<i_{*}<90^{\circ}$, we build an interpolative grid which performs the inverse mapping, i.e., from a critical curve shape to a spin value. The approximation is to use the inferred interferometric shape of the first photon ring as input to this grid as a proxy for a critical curve shape.

Given that the baselines where the interferometric signal is dominated by the $n=1$ photon ring depend on the astrophysical emission profile—and given the lack of detailed knowledge about the astrophysical properties of targeted sources—we adopt an astrophysics-agnostic approach. This involves testing our spin inference scheme across a wide range of astrophysical models, spanning images with very thin photon rings to those with significantly thicker rings. By surveying a broad parameter space, we aim to account for the diversity of potential emission profiles around real astrophysical sources.

In addition to varying the astrophysical models, we also consider several baseline windows in our analysis, demonstrating that our inference method is not only agnostic to the astrophysical assumptions but also to variations in observational configurations, such as receiver frequency and the orbit of the space-based interferometric array. The overarching goal of this proof-of-concept study is to evaluate the robustness of the particular geometric method used in this paper across diverse astrophysical and observational scenarios, while identifying its limitations and range of applicability.

We first consider noiseless time-averaged images of black holes with underlying spin $a_{*}\in\{0.5, 0.94\}$ and low inclination $i_{*}=17^{\circ}$---this particular choice of inclination is motivated by observations of M$87^{*}$ (see, e.g., Ref.~\cite{Mertens_2016}). We apply our spin inference procedure using least-squares-inferred photon ring shapes from interferometric signals spanning $140$ profiles of equatorial emission. We show that for the high spin case, $a_{*}=0.94$, the critical curve shape approximation is sufficiently accurate at (very) long baselines ($u\sim{300}\,\mathrm{G}\lambda
$) to provide a distribution of inferred spins centered on the true value $a_{*}$ with a $2\sigma$ error interval of total width given by $5\%$ of $a_{*}$, where, in this case, the error is quantifying the uncertainty introduced by varying the model of emission, rather than the error of a measurement. At short baselines $(u\lesssim{50}\,\mathrm{G}\lambda)$, the peak of the inferred spin distribution shifts to $5\%$ away from the true value, with a $2\sigma$ interval approximately three times larger than the long baseline case. For $a_{*}=0.5$, the slow rotation of the BH necessitates an approximation of the critical curve shape more accurate than that provided by our inference scheme even at baselines where the $n=2$ photon ring typically dominates, causing the accuracy and precision of our inference scheme to substantially degrade (even in the absence of noise and measurement uncertainty).

We then employ a more statistically robust version of our inference method to explicitly take into account uncertainties in interferometric signals. We restrict this analysis to the high spin case, $a_{*}=0.94$, and consider two canonical emission models. We find that the performance of the inference doesn't significantly change when introducing \textit{only} instrument uncertainty in simulated visibility data. We also apply the method to a simulated dataset of interferometric measurements using \texttt{ngEHTsim}~\cite{2024ApJ...968...69P}, which accounts for weather, instrumental effects, the space-based antenna size, satellite orbit, number and positions of ground telescopes, and the observation duration. This type of synthetic data serves as a proxy for future space-based VLBI missions, such as the proposed Black Hole Explorer (BHEX) mission~\cite{Johnson:2024ttr}. When applying this geometric model to these synthetic observations, we find that the method breaks down, providing only lower bounds of $0.59$ and $0.72$ on the black hole spin at the $3\sigma$ confidence level for the two canonical emission models.

Beyond photon ring asymmetry, as the one studied in this work, several complementary techniques have been proposed for inferring black hole spin (see, e.g., Ref.~\cite{Ayzenberg:2023hfw} for a review). One approach is to measure the slight displacement of the ``inner shadow'' (the lensed horizon image) relative to the photon ring’s critical curve, a geometric offset that grows with spin and inclination~\cite{Chael:2021rjo}. Another class of methods exploits time-domain variability (``hotspots'' or flares): by tracking an orbiting emission feature and correlating its direct and lensed images, one might extract the spin largely independently of uncertain accretion physics~\cite{Tiede:2020jgo}. Similarly, the autocorrelation of stochastic brightness fluctuations on the ring reveals characteristic time delays and angular separations set by geodesics, providing an observable imprint of spin~\cite{Hadar:2020fda,ZhenyuZhang:2025cqn}. Polarimetric signatures have also been identified---for instance, twisted polarization patterns (spirals) in the direct versus secondary ring images change handedness or become more radial at higher spins, offering an independent spin diagnostic~\cite{Himwich_2020,Emami:2022kci,Palumbo:2024czv}. More broadly, the spin is imprinted in the displacement of sub-images from the unlensed position of the black hole on the sky; higher-order images are displaced further, indicating promising astrometric pathways to spin measurement (see, e.g., Ref.~\cite{Johannsen_2010}).

Each of these techniques to measure spin comes with distinct challenges (e.g., demanding stable relative astrometry, long-term monitoring, or high SNR polarimetry). However, together, they leverage different observable consequences of rotation and complement the photon-ring asymmetry approach. As we will show in this work, the asymmetry of the photon ring serves as a morphological indicator of spin, especially for rapidly rotating black holes observed at moderate inclinations. In contrast, the alternative methods harness spatial offsets, temporal correlations, or polarization structures that can isolate spin effects more directly. In concert, these diverse approaches can cross-validate spin measurements, with photon ring asymmetry offering a valuable independent check that relies on different observables than the geometric or time-domain methods. As an example of how these effects may differ in data manifestation, the diameter asymmetry we examine here manifests most plainly in interferometric amplitudes, while relative displacement effects in the image are more natively apparent in interferometric phases, which we do not consider in this work.

Future spaced-based missions, such as BHEX, will measure an abundance of information about the first photon ring: not only the overall size and shape of the ring as a function of angle, but also its angular brightness profile and the displacement of its centroid relative to the $n=0$ image. In contrast, the present analysis utilizes only two observables---the minimum and maximum orthogonal diameters---to probe the asymmetry of the photon ring and infer spin.

The remainder of this paper is organized as follows. Section~\ref{Sec:TheoreticalFramework} provides a summary of the two methods employed in this paper to infer photon ring shapes and the geometric model used for spin inference. In Sec.~\ref{sec:AstroStudy}, we carry out a survey of the inference scheme over $140$ emission models and two BH spins. We then consider the performance of the scheme in the presence of noise and measurement uncertainty in Sec.~\ref{sec:NoiseStudy}. We summarize our results in Sec.~\ref{Sec:Discussion} and provide a discussion of possible improvements of our inference scheme. In Appendix~\ref{App:SpinInclinationInference}, we present another variant of our inference scheme, more suited to Sgr A$^{*}$-like observations, which can be used to infer both spin and inclination, assuming the BH mass is known. 

Throughout, we use an asterisk to denote the true value of the geometric parameters (i.e., spin and inclination) corresponding to those specified in the underlying simulations. The absence of an asterisk denotes either an inferred parameter, or an assumed value of the observer inclination used for spin inference.

\section{Theoretical Framework}
\label{Sec:TheoreticalFramework}

We begin this section by reviewing the methods we employ for inferring photon ring shapes from simulated black hole interferometric signals. We then describe the interpolated grid built from critical curve shapes that we use for spin inference.

\subsection{A Phenomenological Source Model}
\label{ssec:SourceModel}

We simulate black hole images using the Adaptive Analytical Raytracing (\texttt{AART}) code~\cite{CardenasAvendano_2022}. We outline the key ingredients of the model and refer readers to Ref.~\cite{CardenasAvendano_2024} for a comprehensive summary of the phenomenological source model used in our image simulations, and to Ref.~\cite{CardenasAvendano_2022} for the original implementation.

We model an equatorial disk and with an emissivity parameterized by Johnson's standard unbounded (SU) distribution,
\begin{align}
	\label{eq:JonhnsonSU}
	J_{\rm SU}(r;\mu,\vartheta,\gamma)\equiv\frac{e^{-\frac{1}{2}\br{\gamma+\arcsinh\pa{\frac{r-\mu}{\vartheta}}}^2}}{\sqrt{\pa{r-\mu}^2+\vartheta^2}},
\end{align}
where the parameters, $\mu$, $\vartheta$, and $\gamma$ control the location, width and asymmetry, respectively, of the emission's peak. This radial profile should be interpreted as a proxy for a time-averaged black hole image, representing the source intensity, denoted as $ I_{\rm s} $. The observed intensity, $ I_{\rm o} $, at each pixel in the image, $(\alpha,\beta)$, is the sum of contributions from all $ n $ photon rings that contribute to that pixel. It is computed as $ I_{\rm o} = \zeta_{n} g_{n}^3 I_{\rm s} $, where $ \zeta $ is a ``fudge'' factor, aimed to account for neglected thickness effects, and $ g $ represents the observed redshift~\cite{CardenasAvendano_2022}.

Given a black hole image, we compute its complex radio visibility, 
\begin{align}
	\label{eq:ComplexVisibility}
	V(\mathbf{u})=\int I_{\rm o}(\mathbf{x}_{\rm o})e^{-2\pi i\mathbf{u}\cdot\mathbf{x}_{\rm o}}\ed^2\mathbf{x}_{\rm o},
\end{align}
where $\mathbf{x}_{\mathrm{o}}$ denotes dimensionless sky coordinates of a distant observer at radius $r_{\mathrm{o}}\gg M$, and $\mathbf{u}$ denotes the dimensionless vector of separation between telescopes in the array, projected onto the plane perpendicular to the line of sight and measured in units of the observational wavelength. Following Refs.~\cite{GLM2020, CardenasAvendano_2022}, $V(\mathbf{u})$ is computed not via the $2$D Fourier transform (\ref{eq:ComplexVisibility}), but by applying the projection-slice theorem to obtain the visibility amplitude $|V(u, \varphi)|$ across fixed slices at polar angle $\varphi$ in the Fourier plane, which we normalize as $|V(0)|=0.6\,\mathrm{Jy}$ since we are interested in modeling observations of M$87^*$~\cite{EventHorizonTelescope:2019ths}. We also assume a mass-to-distance ratio $(M/r_{\rm o})_{\rm M87^*}=3.62\,\mu\mathrm{as}$, and, for all the images used in this work, a pixel resolution of $0.05\,M$ and an observer screen of size $[-25\,M,25\,M]$.

\subsection{Photon Ring Shape Inference}
\label{ssec:ShapeInference}

In the previous section, we briefly discussed the source model's main ingredients and our main observable: the visibility amplitude. We now describe two inference methods used to analyze the simulated images in visibility space: one which is computationally efficient and thus suitable for analyzing a large number of images, and another which is more statistically robust but also more computationally expensive.

\subsubsection{Inference Method I: Least Squares Curve Fitting}
\label{sssec:LeastSquares}

For a given image, there exists sufficiently long baselines $u$ where $1/d \ll u \ll 1/w$---here, $d$ and $w$  are the diameter and width, respectively, of a ring. In this ``universal regime,'' the visibility amplitude takes on the form~\cite{Gralla_2020}
\begin{align}
	\label{eq:UniversalVisibility}
    |V(u,\varphi)|&=\frac{\sqrt{\pa{\alpha_\varphi^{\rm L}}^{2}+\pa{\alpha_\varphi^{\rm R}}^{2}+2\alpha_\varphi^{\rm L}\alpha_\varphi^{\rm R}\sin\pa{2\pi d_\varphi u}}}{\sqrt{u}},
\end{align}
where $\alpha_\varphi^{\rm L/R}$ encode the intensity of the ring, and $d_{\varphi}$ denotes its projected image-plane diameter. We extract the projected diameters $d_\varphi$ from the visibility amplitude across some baseline window $u\in[u_{1},u_{2}]\,\mathrm{G}\lambda$ using \texttt{SciPy}'s \verb|curve_fit| to perform a nonlinear least squares fit to the functional form 

\begin{align}
	\label{eq:VisibilityFit}
    V_{\mathrm{fit}}(u; d_{\varphi}, \alpha_\varphi^{\rm L}, \alpha_\varphi^{\rm R}, p) = e^{-pu}\,|V(u,\varphi)|,
\end{align}
where we have included an additional parameter, $p$, to account for (exponential) decay in the signal at short baselines ($u\lesssim 100\,\mathrm{G}\lambda$)~\cite{Johnson_2020}. At baselines where the $n=1$ ring dominates, the $d_{\varphi}$ no longer describe image-plane ring diameters, which are ill-defined for the first photon ring due to its non-negligible width, but instead constitute visibility-space diameters which describe the interferometric ringing of the signal~\cite{CardenasAvendano_2023}.

As in Ref.~\cite{CardenasAvendano_2024}, we carry out visibility amplitude fits to obtain diameters $d_{\varphi}$ for $36$ equally-spaced Fourier plane angles $\varphi\in\{0^{\circ}, 5^{\circ},\ldots,175^{\circ}\}$. For each slice, the fitted $d_{\varphi}$ is not guaranteed to be a global minimum of the root-mean-square deviation (RMSD), so we consider additional candidate ring diameters which are local minima of the RMSD. This generates a family of curves called ``circlipses'', which, for a photon ring, has the GR-predicted functional form~\cite{GrallaLupsasca2020c,Paugnat2022}:
\begin{align}
	\label{eq:Circlipse}
	\frac{d_\varphi}{2}
    &=R_{0}+\sqrt{R_1^2\sin^2\pa{\varphi-\varphi_0}+R_2^2\cos^2\pa{\varphi-\varphi_0}},
\end{align}
which represents the sum of a circle of radius $R_0$ and an ellipse with axes $R_1$ and $R_2$, while the additional phase parameter $\varphi_0$ accommodates potential rotations of the image. 

From the family of candidate circlipses, the ``best'' circlipse
is identified as the one with the lowest total RMSD summed over its constituent diameters. We fit the best circlipse to the functional form~\eqref{eq:Circlipse}, obtaining the best-fit parameters $R_0,R_{1},$ and $R_2$, from which we compute the maximal, orthogonal ring diameters
\begin{align}
	d_\parallel&=2(R_0+R_1),\\
    d_\perp&=2(R_0+R_2),    
\end{align}
Together, these diameters define the fractional shape asymmetry of the ring:
\begin{align}
	\label{eq:FractionalAsymmetry}
    f_A=1-\frac{d_\perp}{d_\parallel}=\frac{R_{1}-R_{2}}{R_{0}+R_{1}}.
\end{align}

As explained in Ref.~\cite{CardenasAvendano_2024}, not all fits attempted using this method are successful, and some of those that are can be of a low quality. We maintain the threshold of $\mathrm{RMSD}<0.05\%$, chosen empirically, for a circlipse fit to be deemed ``successful'' (and subsequently used for spin inference). 

\subsubsection{Inference Method II: Markov Chain Monte Carlo Simulations}
\label{sssec:MCMC}

The advantage of the fitting method described in the previous section is that it is sufficiently fast (taking $\sim1$\,s to obtain an inferred circlipse) to carry out circlipse inferences across the several emission profiles and baseline windows studied in this paper, allowing us to understand how varying the astrophysics affects the performance of our spin inference procedure. However, we are also interested in the robustness of our inference procedure against noise and measurement uncertainties, whose effects we study by fixing the emission profile and using a series of Markov Chain Monte Carlo (MCMC) simulations to carefully propagate errors into the final spin estimate.

For a given model of the emission, we first run MCMC simulations on the $36$ simulated visibility amplitudes to infer $36$ sets of model parameters $\mathbf{\theta}_{\mathrm{vis}}=(d_{\varphi}, \alpha^{\mathrm{L}}_{\varphi}, \alpha^{\mathrm{R}}_{\varphi}, p_\varphi)$ from the likelihood

\begin{align}
    \mathcal{L}_{\mathrm{vis}}(\bm{\theta_{\mathrm{vis}}})&=\prod_{i}\frac{1}{\sqrt{2\pi\sigma_{i}^2}}\,\exp\left(-\frac{(V_{i} - m_{i}(\bm{\theta_{\mathrm{vis}}}))^{2}}{2\sigma_{i}^2}\right),\label{eq:VisibilityLikelihood}
\end{align}
where $V_{i}$ and $\sigma_{i}$ are the visibility amplitude data points and $1\sigma$ errors for a given cut and the $m_{i}(\theta_\mathrm{vis})$ are model predictions using Eq.~\ref{eq:VisibilityFit}.

If the posterior distributions of the diameters $d_\varphi$ are all single-peaked, there is only one inferred circlipse which we proceed to fit to the functional form (\ref{eq:Circlipse}). Depending on the baseline window, however, the $36$ posterior distributions of the $d_{\varphi}$ may be multi-peaked~\cite{CardenasAvendano_2024}. In this case, the procedure for determining the ``best'' circlipse proceeds similarly to that described in Sec.~\ref{sssec:LeastSquares}. Therein, the candidate diameters were the local minima of the RMSD, whereas, in the MCMC approach, the candidate diameters arise from secondary peaks in the posterior distributions of the $d_{\varphi}$. For each posterior distribution, we fit each peak therein to a Gaussian functional form to obtain its central value and width, and compute the associated posterior probability from the total area under the fitted curve. The result is a family of candidate circlipses, from which we identify the best circlipse as that with the largest product of posterior probabilities over its constituent diameters.

Once the diameters and widths $(d_{\varphi}, \sigma_{\varphi})$ of the best circlipse have been obtained, we infer the circlipse parameters $\mathbf{\theta}_{\mathrm{circ}}=(R_{0}, R_{1}, R_{2}, \varphi_{0})$ via another MCMC simulation with likelihood
\begin{align}
    \mathcal{L}_{\mathrm{circ}}(\bm{\theta}_{\mathrm{circ}})&=\prod^{175^{\circ}}_{\varphi=0^{\circ}}\frac{1}{\sqrt{2\pi\sigma_{\varphi}^2}}\,\exp\left(-\frac{(d_{\varphi} - m_{\varphi}(\bm{\theta}_{\mathrm{circ}}))^{2}}{2\sigma_{\varphi}^2}\right),\label{eq:CirclipseLikelihood}
\end{align}
where $m_{\varphi}(\mathbf{\theta}_\mathrm{circ})$ are model predictions using Eq.~\ref{eq:Circlipse}. We then compute the posterior distribution of the fractional asymmetry by pointwise evaluation of Eq.~\ref{eq:FractionalAsymmetry} using the sampling chains associated with the posterior distributions of $R_{0}$, $R_{1}$, and $R_{2}$. Following this procedure, we always obtain a single-peaked posterior distribution of the fractional asymmetry, which we again fit to a Gaussian to obtain the location of its peak and its width.

From the inferred confidence interval for the fractional asymmetry, we infer a central value of spin and its associated error using the predictive grid described in the next section.

\subsection{The Geometric Model: The Shape of the Critical Curve}
\label{ssec:ShapeOfCritCurve}

Since the critical curve~\cite{Bardeen1973} depends only on the geometric parameters (black hole spin and inclination), we use it as a proxy to infer these parameters. In the observer's sky, this (theoretical) curve represents the apparent image of photon orbits that are asymptotically bound. On the observer's screen, $(\alpha,\beta)$, it forms a closed, convex curve with an angular-dependent diameter $d_{\varphi}$.

As in Ref.~\cite{Johnson_2020}, we characterize the ``shape'' of the critical curve by the diameter-asymmetry pair $(d_{||}, f_{A})$, where $d_{||}$ denotes the maximum image-plane diameter parallel to the spin axis of the black hole projected onto the observer screen, e.g., $d_{||}=d_{\varphi=90^{\circ}}$ (if the image is not rotated), $d_{\perp}=d_{\varphi=0^{\circ}}$ is the diameter perpendicular to the projected spin axis, and $f_{A}\equiv 1 - d_{\perp}/d_{||}$ is the fractional shape asymmetry. 

\begin{figure}
    \centering
    \includegraphics[width=\columnwidth]{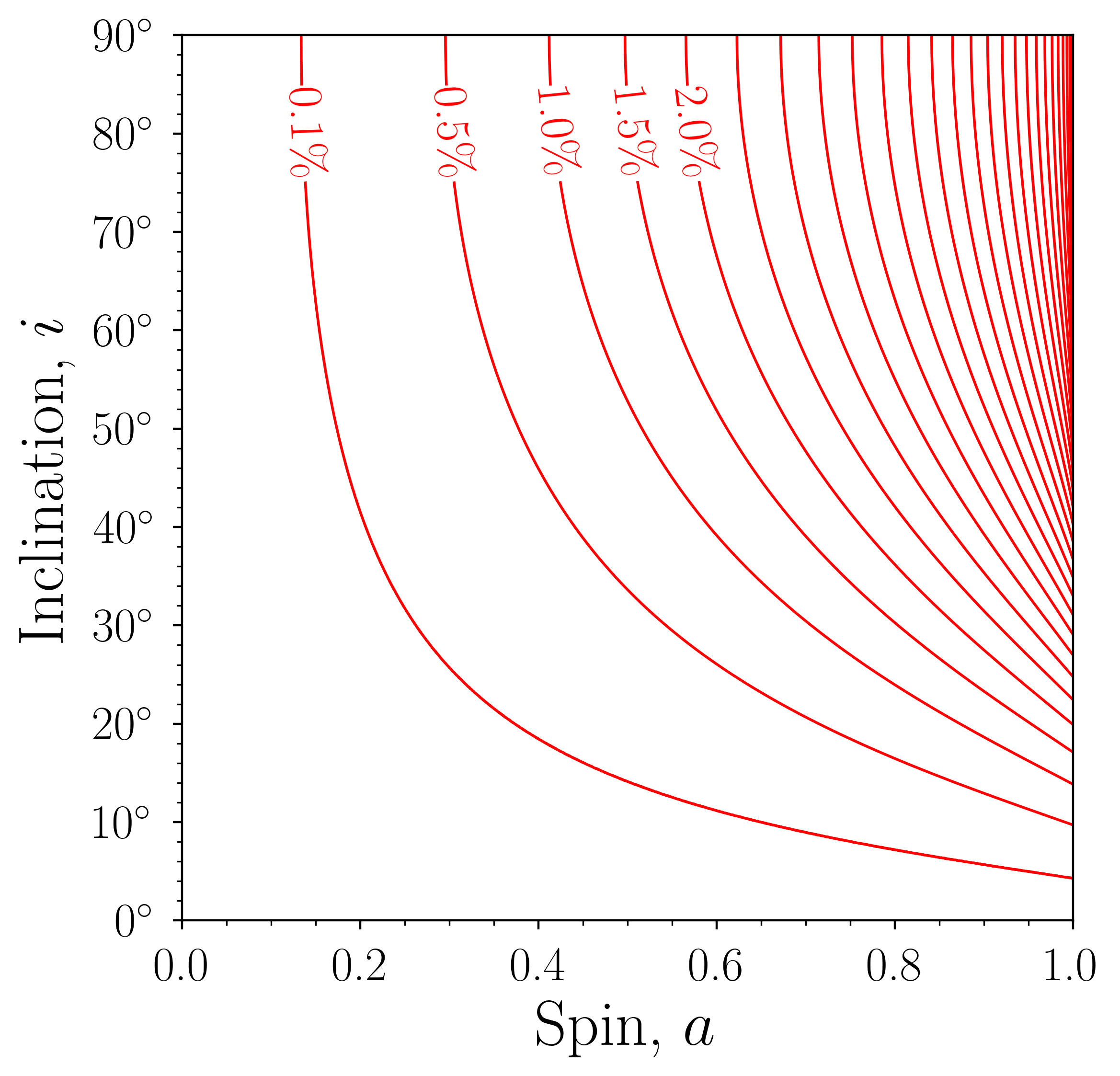}
    \caption{Fractional shape asymmetry (\ref{eq:FractionalAsymmetry}) contours interpolated from critical curves with spin-inclination pairs $(a_{*},i_{*})$, where $a_{*}$ and $i_{*}$ take on $2,000$ equally-spaced values in the ranges $[0.00001, 0.999999]$, and $[1^{\circ}, 90^{\circ}]$, respectively. The first (leftmost) contour corresponds to a critical curve fractional asymmetry of $0.1\%$ while the respective asymmetry of the subsequent contours follow the arithmetic sequence $0.5\%$, $1.0\%,\ldots,13.0\%$. The $0.0\%$ asymmetry contour lies on the $y$-axis, since the critical curve is circular for $a_{*}=0$ at all inclinations $0^{\circ}\leq i\leq90^{\circ}$.}
    \label{fig:AsymmetryContours}
\end{figure}

We compute the shape of the critical curve for spin-inclination pairs $(a_{*}, i_{*})$ where $a_{*}$ and $i_{*}$ take on $2000$ equally-spaced values in the ranges $[0.00001, 0.999999]$, and $[1^{\circ}, 90^{\circ}]$, respectively. For each spin-inclination pair, we compute the corresponding critical curve, its diameters $d_{\perp}$ and $d_{||}$, and its fractional asymmetry, which, over the aforementioned parameter space, takes on values between $0\%$ and $13.4\%$. We use this data to build an interpolated grid of fractional asymmetry as a function of spin and inclination as shown in Fig.~\ref{fig:AsymmetryContours}, wherein a specified shape asymmetry has a corresponding contour in the spin-inclination plane. 

With this underlying theoretical model, our spin inference procedure consists of first inferring a photon ring fractional asymmetry across some baseline window using either method described in Sec.~\ref{ssec:ShapeInference}. This inferred asymmetry is then used as input to the grid depicted in Fig.~\ref{fig:AsymmetryContours} (and thus as an approximation of underlying critical curve's shape asymmetry), defining some contour in the spin-inclination plane. The estimated spin is then given by the $x$-value of the intersection between the asymmetry contour and the horizontal line defined by some assumed value of the observer inclination $i$. If there is no such intersection, this method fails to provide a sharp prediction of the spin.

When performing spin inference in this paper, we will assume observer inclinations $i\in\{15^{\circ}, 17^{\circ}, 19^{\circ}\}$, i.e., sampling around the currently favored and tightly constrained viewing inclination of $17^\circ$~\citep{Mertens_2016}. Unless otherwise stated, all other confidence intervals stated in this paper are at the $2\sigma$ level.

Using the method described in Sec.~\ref{ssec:ShapeInference}, the inferred diameters $d_{\varphi}$ are in units of microarcseconds, while the critical curve diameters we compute are in units of the BH mass. By depending only on the ratio $d_{\perp}/d_{||}$, the fractional asymmetry is dimensionless, allowing us to carry out spin inference without an additional assumption about the mass of the underlying black hole. 

For sources like Sgr A$^{*}$, where the mass is well constrained but not the inclination, an alternative version of this inference procedure could be used. By assuming the mass, one can convert inferred diameters into units of the BH mass, overlay contours of both the inferred shape asymmetry and the parallel diameter $d_{||}$ in the spin-inclination plane (e.g., Fig.~\ref{fig:DiameterAsymmetryContours}), and infer both spin and inclination by computing the $x$ and $y$ values of the intersection between these two contours. We present some results using this approach in App.~\ref{App:SpinInclinationInference}, where we  show that the inferred diameters $d_{\varphi}$ are far less robust against changing the astrophysics (even in the absence of noise) compared to the fractional asymmetry, causing the inferred spin to vary significantly depending on the choice of emission profile at short baselines. 

Thus far, we have outlined our methods for simulating black hole images and their corresponding visibility amplitudes, inferring a photon ring shape, and estimating the black hole's spin. In the remainder of this paper, we apply this spin inference scheme to time-averaged black hole images without noise and then to simulated observations.

\section{Survey over astrophysical emission profiles}
\label{sec:AstroStudy}

We first quantify the effect of varying the underlying (astrophysical) model of emission on the performance of our spin inference method. We attempt spin inference for black holes with underlying inclination $i_{*}=17^{\circ}$, spins $a_{*}\in\left\{0.5,0.94\right\}$, and $140$ equatorial emission profiles as specified in Tab.~\ref{tbl:Parameters}.

\begin{table}
	\centering
	\begin{tabular}{c c}
 	\hline
    \hline
	Johnson SU Parameter & Values \\
	\hline
	$\mu$ & $r_-,r_+/2,r_+,3r_+/2,2r_+$ \\
	$\gamma$ & $-2,-1.5, -1,0,1,1.5, 2$ \\
	$\vartheta/M$ & $0.25, 0.5, 1.0, 1.5$ \\
 	\hline
    \hline
	\end{tabular}
	\caption{Johnson's SU parameters for the $140$ emission models considered in this paper, as shown in Fig.~\ref{fig:EmissionProfiles}. The parameter $\mu$ controls the location of the peak of the emission, $\vartheta$ the width of the peak, and $\gamma$ its asymmetry.
	The outer/inner event horizon radii are given by $r_\pm=M\pm\sqrt{M^2-a^2}$.}
	\label{tbl:Parameters}
\end{table}

Using \texttt{AART}, we simulate the interferometric signature of a black hole with spin $a_{*}$, inclination $i_{*}$, and equatorial emission profile $(\mu, \gamma, \vartheta)$ belonging to the parameter space specified in Tab.~\ref{tbl:Parameters}. Using the least-squares inference scheme described in Sec.~\ref{sssec:LeastSquares}, we infer from the interferometric signal a fractional shape asymmetry across four baseline windows: $[20, 40]\,\mathrm{G}\lambda$, $[30, 50]\,\mathrm{G}\lambda$,  $[80, 100]\,\mathrm{G}\lambda$,  $[280, 300]\,\mathrm{G}\lambda$. These windows were selected to cover observations that range from a baseline window near the $n=0$ to $n=1$ transition to a baseline window primarily dominated by $n=2$ for most astrophysical profiles. For instance, at $230$ GHz, a baseline of $40$ G$\lambda$ corresponds to a telescope separation of approximately $4$ Earth diameters. We then attempt spin inference by using the inferred asymmetry as input to the grid depicted in Fig.~\ref{fig:AsymmetryContours}, assuming we know the underlying inclination $i_{*}=17^{\circ}$ exactly.

\begin{figure*}
    \centering
    \includegraphics[width=\textwidth]{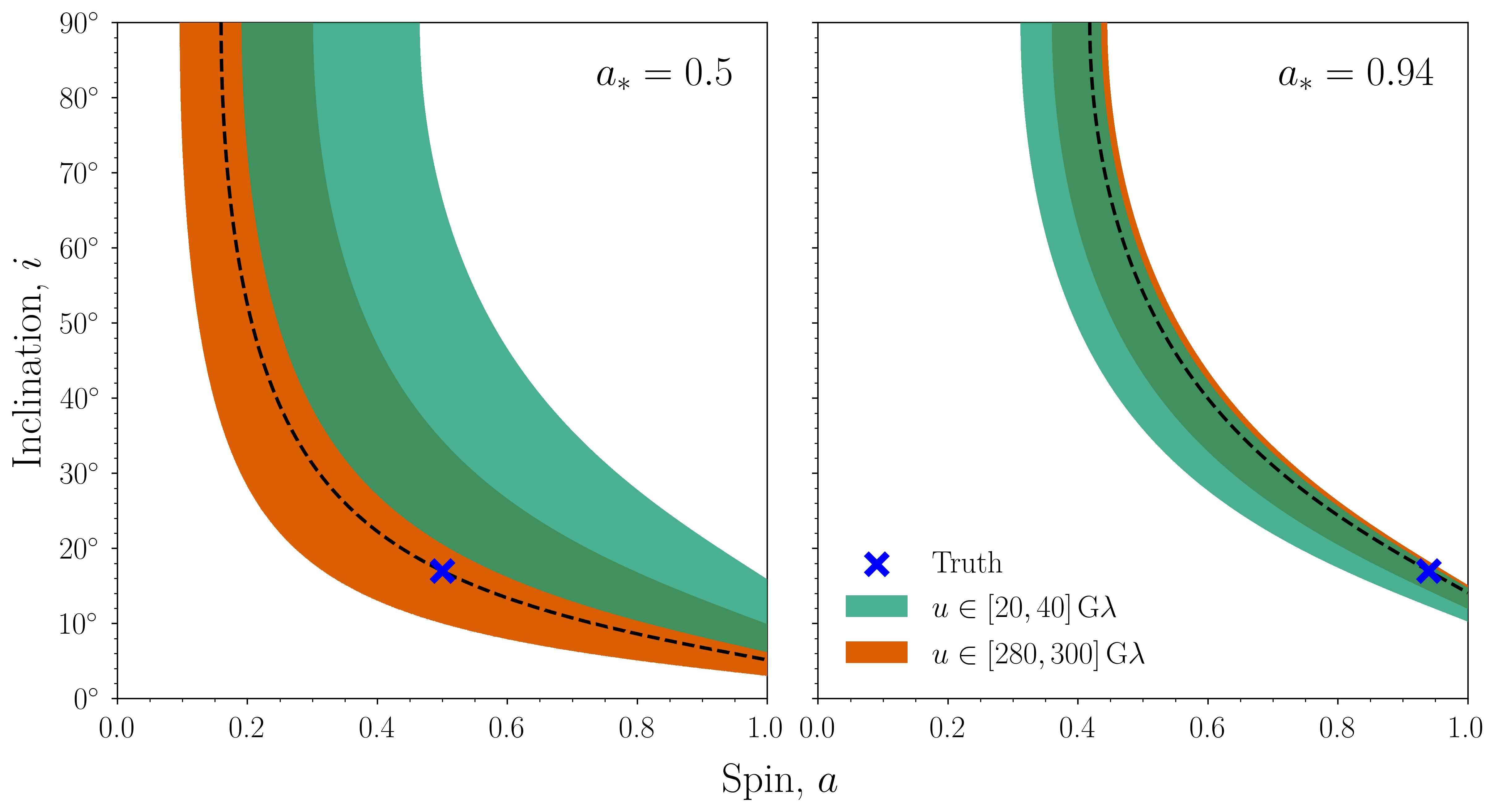}
    \caption{Contour bands in the spin-inclination plane for the photon ring shape asymmetries inferred across the baseline windows $[20, 40]\,\mathrm{G}\lambda$ (green and where the $n=1$ photon ring typically dominates the signal) and $[280, 300]\,\mathrm{G}\lambda$ (orange and where the $n=2$ photon ring typically dominates the signal) with underlying spins $a_{*}=0.5$ (left) and $a_{*}=0.94$ (right). Each band indicates the range of inferred asymmetries: the left and right edge of each band is given by the minimum and maximum inferred asymmetry, respectively, while the region in between is filled in for visual clarity. Here (and in subsequent figures) ``Truth'' denotes the point $(a_{*}, i_{*})$. The black dashed lines denote the fractional asymmetry of the underlying critical curve circlipses---by construction, the closer any inferred asymmetry is to this value, the more accurate the spin inference. For the underlying spin $a_{*}=0.5$, the critical curve asymmetry is $0.13\%$ and the minimum and maximum inferred asymmetries are $(0.21\%,1.30\%)$ across $[20, 40]\,\mathrm{G}\lambda$, and $(0.05\%,0.52\%)$ across $[280, 300]\,\mathrm{G}\lambda$. For $a_{*}=0.94$, the critical curve asymmetry is $1.02\%$, and the minimum and maximum inferred asymmetries are $(0.56\%,1.13\%)$ across $[20, 40]\,\mathrm{G}\lambda$, and $(0.75\%,1.18\%)$ across $[280, 300]\,\mathrm{G}\lambda$.}
    \label{fig:AsymmetryBands}
\end{figure*}

\begin{figure*}
    \centering
    \includegraphics[width=\textwidth]{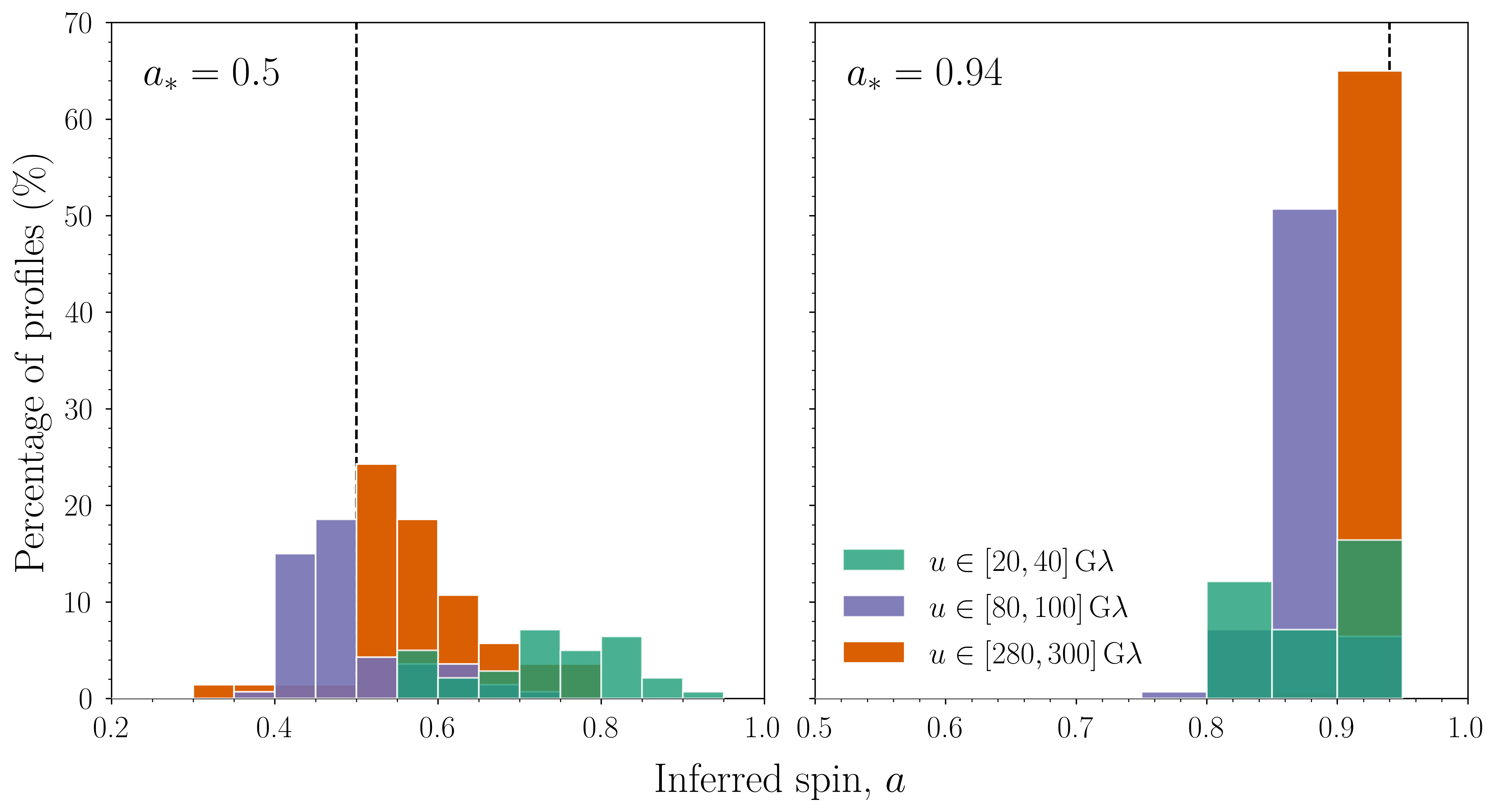}
    \caption{Distributions of the spin inferred across the baseline windows $[20, 40]\,\mathrm{G}\lambda$ (green), $[80, 100]\,\mathrm{G}\lambda$ (purple),  and $[280, 300]\,\mathrm{G}\lambda$ (orange) with underlying spins $a_{*}=0.5$ (left) and $a_{*}=0.94$ (right). Each histogram has a corresponding set of inferred asymmetries (which lie within bands like those depicted in Fig.~\ref{fig:AsymmetryBands}) from which the spin distribution is obtained by finding the point of intersection of each asymmetry contour with the horizontal line $i=17^{\circ}$ in the spin-inclination plane. Estimated $2\sigma$ error intervals for each distribution are listed in Tab.~\ref{tbl:AstroSpinInferenceStatistics}.}
\label{fig:SpinDistribution}
\end{figure*}

We show in Fig.~\ref{fig:AsymmetryBands} the range of inferred shape asymmetries over all the emission profiles across the baseline windows $[20, 40]\,\mathrm{G}\lambda$ and  $[280, 300]\,\mathrm{G}\lambda$, and the critical curve asymmetry for underlying spins $a_{*}=0.5$ (left panel) and $a_{*}=0.94$ (right panel). Each band represents only the range of inferred asymmetries---their distribution consists of a discrete set of contours which lie completely within a given band. We then obtain the distributions of inferred spins shown in Fig.~\ref{fig:SpinDistribution}, and whose summary statistics are listed in Tab.~\ref{tbl:AstroSpinInferenceStatistics}, by finding the intersection of each contour in a given band with the horizontal line $i=17^{\circ}$.

For an underlying spin of $a_{*}=0.5$, we find that although increasing the baseline tends to make the spin inference more accurate, i.e., the central value of the inferred spin distributions move towards the true value as shown in Fig.~\ref{fig:SpinDistribution}, there is not an accompanying decrease in sensitivity to the choice of emission profile since the width of the distributions remain comparably large. Across the window $[20, 40]\,\mathrm{G}\lambda$, the spin inference is both inaccurate and imprecise: the central value of the inferred spin distribution is $\sim{50}\%$ larger than the true value with a $2\sigma$ width given by $70\%$ of the true spin value. Across the longer-baseline window $[280, 300]\,\mathrm{G}\lambda$, the method is similarly imprecise but somewhat more accurate, with the center of the distribution $14\%$ larger than the true spin value.

The breakdown of our spin inference method in this case can be attributed to the increased degeneracy of the critical curve shape at lower values of the spin parameter (with fixed $i_{*}=17^{\circ}$) coupled with less accurate ring shape inferences. As the spin of the underlying black hole decreases to zero, the critical curve circularizes and its shape asymmetry changes less for fixed increments of spin. Consequently, for lower values of $a_{*}$, one requires an increasingly accurate approximation of the critical curve shape to infer spin within a fixed level of accuracy using this method---this can also be seen in Fig.~\ref{fig:AsymmetryContours}, wherein fixed intervals of asymmetry have corresponding contours which are increasingly separated at low spins. Meanwhile, at this value of spin, our shape inference scheme is both less accurate and less precise.
For example, in the case $a_{*}=0.5$ and $i_{*}=17^{\circ}$, the critical curve asymmetry is $0.13\%$, while, across the range of emission profiles here considered, we infer a range of shape asymmetries $f_{A}\in[0.21\%, 1.30\%]$ across the baseline window $[20, 40]\,\mathrm{G}\lambda$, which doesn’t include the critical curve asymmetry and which is more than twice as large as the range inferred for $a_{*}=0.94$ across the same baseline window (which does include the corresponding critical curve asymmetry). 

\begin{table}
\resizebox{\columnwidth}{!}{
\renewcommand{\arraystretch}{2.5}
\begin{tabular}{ccccc}
\cline{2-5} \cline{3-5} \cline{4-5} \cline{5-5} 
\multicolumn{1}{c|}{} & \multicolumn{2}{c|}{$a_{*}=0.5$} & \multicolumn{2}{c|}{$a_{*}=0.94$}\tabularnewline
\hline 
\multicolumn{1}{|c|}{\shortstack{Baseline\\Window}} & \multicolumn{1}{c|}{\shortstack{Inferred\\Spin}} & \multicolumn{1}{c|}{\shortstack{Successful\\ Profiles}} & \multicolumn{1}{c|}{\shortstack{Inferred\\Spin}} & \multicolumn{1}{c|}{\shortstack{Successful\\Profiles}}\tabularnewline
\hline
\hline
$[20, 40]\,\mathrm{G}\lambda$ & $0.75^{+0.19}_{-0.16}$ & $33\%$ & $0.89^{+0.06}_{-0.07}$ & $38\%$ \tabularnewline
$[30, 50]\,\mathrm{G}\lambda$ & $0.65^{+0.23}_{-0.17}$ & $61\%$ & $0.89^{+0.06}_{-0.08}$ & $59\%$ \tabularnewline
$[80, 100]\,\mathrm{G}\lambda$ & $0.47^{+0.2}_{-0.08}$ & $48\%$ & $0.88^{+0.09}_{-0.07}$ & $69\%$ \tabularnewline
$[280, 300]\,\mathrm{G}\lambda$ & $0.57^{+0.19}_{-0.19}$ & $72\%$ & $0.94^{+0.02}_{-0.03}$  & $69\%$ \tabularnewline
\hline
\hline
\end{tabular}
}\caption{Spin inference results for the survey over the $140$ emission profiles listed in Tab.~\ref{tbl:Parameters} with underlying spins $a_{*}=0.5$ and $a_{*}=0.94$. We use a Gaussian kernel density estimator on the histograms shown in Fig.~\ref{fig:SpinDistribution} to estimate a $2\sigma$ confidence interval for the spin, listed as $\pm$ values around the central $50\%$ spin estimate. The errors quoted here are sourced by varying the model of emission (and not due to measurement uncertainties as considered in Sec.~\ref{sec:NoiseStudy}). We also list the percentage of emission profiles for which we successfully inferred a spin value.}\label{tbl:AstroSpinInferenceStatistics}
\end{table}

For an underlying spin of $a_{*}=0.94$, the inference scheme is more accurate and robust against changes in the emission model. The central value of the spin distribution inferred across the window $[20, 40]\,\mathrm{G}\lambda$ is $\sim5\%$ less than the true value, while, for the window $[280, 300]\,\mathrm{G}\lambda$, the central value matches the true value exactly. We also see that, as expected, the spin inferred across $[280, 300]\,\mathrm{G}\lambda$, where the $n=2$ ring typically dominates, has a narrower $2\sigma$ interval than the $n=1$ dominated windows, indicating that the shape (and hence spin) inferred in this regime is both more accurate and varies less with changes in the emission model.

In the rapidly-rotating case, the spin distributions inferred across the typically $n=1$ dominated windows $[20, 40]\,\mathrm{G}\lambda$, $[30, 50]\,\mathrm{G}\lambda$, and $[80, 100]\,\mathrm{G}\lambda$ do not vary significantly in either their central value or width. This suggests that, using this method, there isn't much advantage gained (on average) in terms of accuracy by increasing the observational baseline. Rather, we see from the values listed in Tab.~\ref{tbl:AstroSpinInferenceStatistics} that the main effect of increasing the baseline in this study is to increase the fraction of the $140$ emission profiles for which we were able to successfully infer spin. This effect appears most pronounced at short baselines: shifting the baseline by only $10\,\mathrm{G}\lambda$ from $[20, 40]\,\mathrm{G}\lambda$ to $[30, 50]\,\mathrm{G}\lambda$ increased the percentage of successful inferences from $38\%$ of the $140$ emission profiles to $59\%$, while shifting the baseline by another $50\,\mathrm{G}\lambda$ only led to a further additive increase  of $10\%$. 

\begin{figure*}
    \centering
    \includegraphics[width=\textwidth]{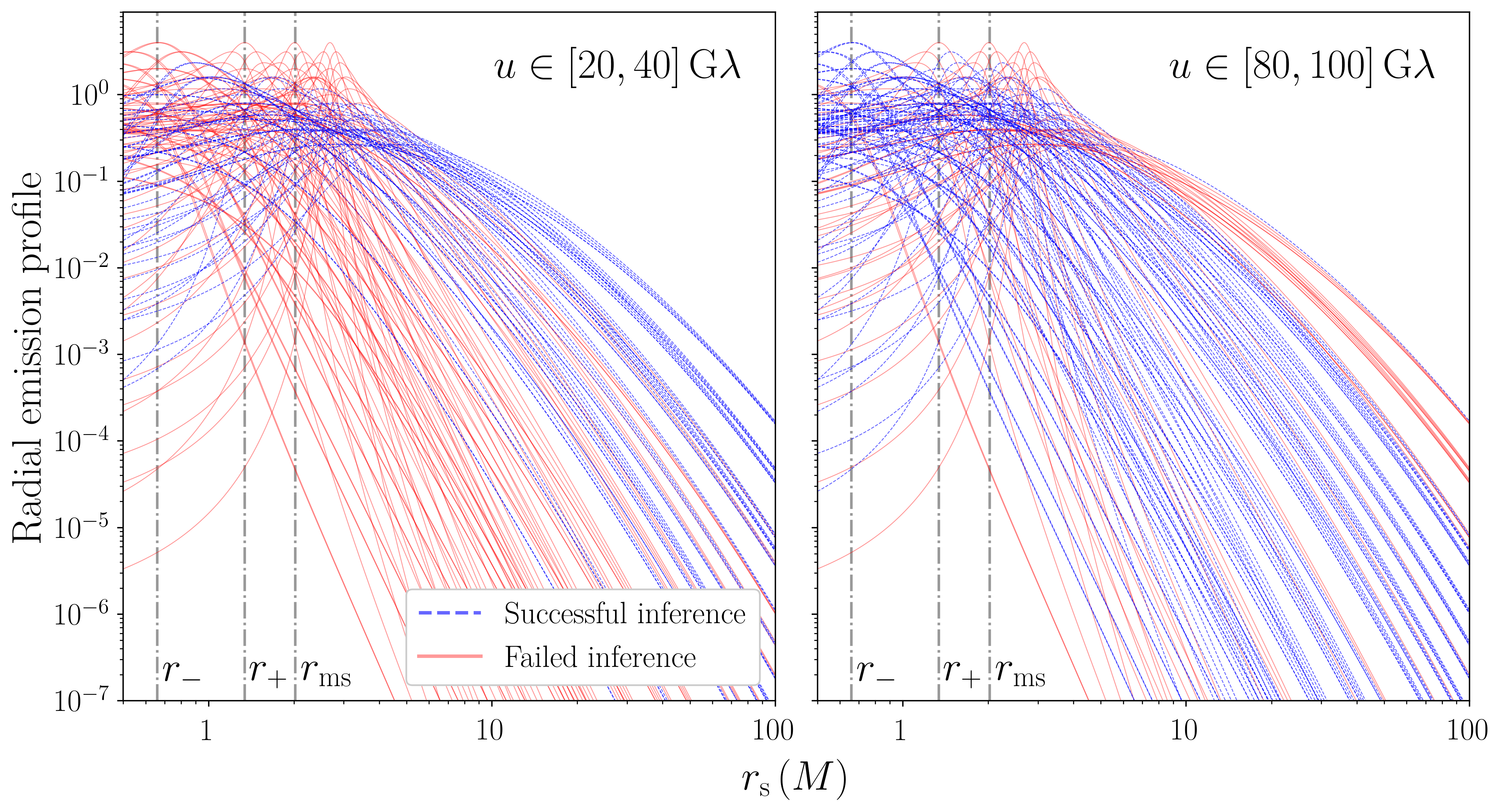}
    \caption{Equatorial $J_{\rm SU}(r;\mu,\vartheta,\gamma)$ emission profiles as a function of the Boyer-Lindquist radius for the $140$ profiles listed in Tab.~\ref{tbl:Parameters}. The blue curves correspond to profiles for which spin inference from the corresponding visibility amplitude across the baseline windows $[20, 40]\,\mathrm{G}\lambda$ (left) and $[80, 100]\,\mathrm{G}\lambda$ (right) was successful, while red denotes unsuccessful inferences. In all cases, the underlying black hole has spin $a_{*}=0.94$ and inclination $i_{*}=17^{\circ}$. The dash-dotted grey lines denote the location of the inner/outer event horizons $r_{\pm}$ and the inner-most stable circle orbit $r_{\mathrm{ms}}$. As listed in Tab.~\ref{tbl:AstroSpinInferenceStatistics}, spin inference was successful across $[20, 40]\,\mathrm{G}\lambda$ for $38\%$ of the emission profiles, compared to $69\%$ across $[80, 100]\,\mathrm{G}\lambda$.}
\label{fig:EmissionProfiles}
\end{figure*}

The sensitivity of the spin inference success rate to the baseline is largely due to regions of transition in the visibility amplitude~\cite{Paugnat2022,CardenasAvendano_2024}, where our photon ring shape inference scheme breaks down. We plot in Fig.~\ref{fig:EmissionProfiles} the $140$ emission profiles listed in Tab.~\ref{tbl:Parameters} as a function of (Boyer-Lindquist) source radius, highlighting in blue the profiles for which spin inference was successful across the windows $[20, 40]\,\mathrm{G}\lambda$ and $[80, 100]\,\mathrm{G}\lambda$, and in red those for which the inference was unsuccessful. Across the window at shorter baselines, the more rapidly decaying profiles tend to be in regions of transition, and, consequently, the inference procedure fails for such profiles. Across the window at longer baselines, the success rate for these profiles increases since they have entered the $n=1$ dominated regime, contributing to the overall increase of $31\%$ in the spin inference success rate between these two windows.

\section{Measurement uncertainty and noise propagation}
\label{sec:NoiseStudy}

We now provide a proof-of-concept example of data analysis techniques that can be used to apply our spin inference method to realistic visibility amplitude data by explicitly accounting for measurement errors. We simulate data from a source with fixed spin, inclination, and emission model, and then attempt spin inference.

In Sec.~\ref{ssec:SpinInferenceGLM}, we consider \texttt{AART}-simulated visibility amplitudes with artificially-introduced measurement errors which are propagated into the spin inference using the MCMC-based method described in Sec.~\ref{sssec:MCMC}. In Sec.~\ref{ssec:SpinInferenceBHEX}, we consider more comprehensive sources of measurement errors by applying the same data analysis techniques to attempt spin inference from simulated visibility data from observational forecasts. 

\subsection{Measurement Uncertainty in Purely \texttt{AART}-simulated Visibilities}
\label{ssec:SpinInferenceGLM}

We fix the underlying black hole spin and inclination to be $a_{*}=0.94$ and $i_{*}=17^{\circ}$, respectively, and the emission profile to have $J_{\mathrm{SU}}$ parameters $\mu=r_{-}$, $\vartheta=0.5\,M$, and $\gamma=-1.5\,M$, for which the simulated visibility profile (shown in Fig.~1 of Ref.~\cite{CardenasAvendano_2024}) is broadly consistent with the EHT results of M$87^{*}$ on Earth baselines~\cite{GLM2020}.

For each baseline cut $\varphi$, we assign to each point of the simulated visibility amplitude a uniform uncertainty given by $33\%$ of the power across the given baseline window. Using a constant fractional error (rather than a constant absolute error) can cause the maximum-likelihood solution to deviate from a unit-weight fit, and such a model does not accurately capture VLBI thermal noise, which is generally source-model independent and can vary with antenna elevation and atmospheric conditions. Nonetheless, this simplified approach provides a tractable way to test our spin-inference framework under a controlled noise prescription. 

\begin{figure*}
    \centering
    \includegraphics[width=\textwidth]{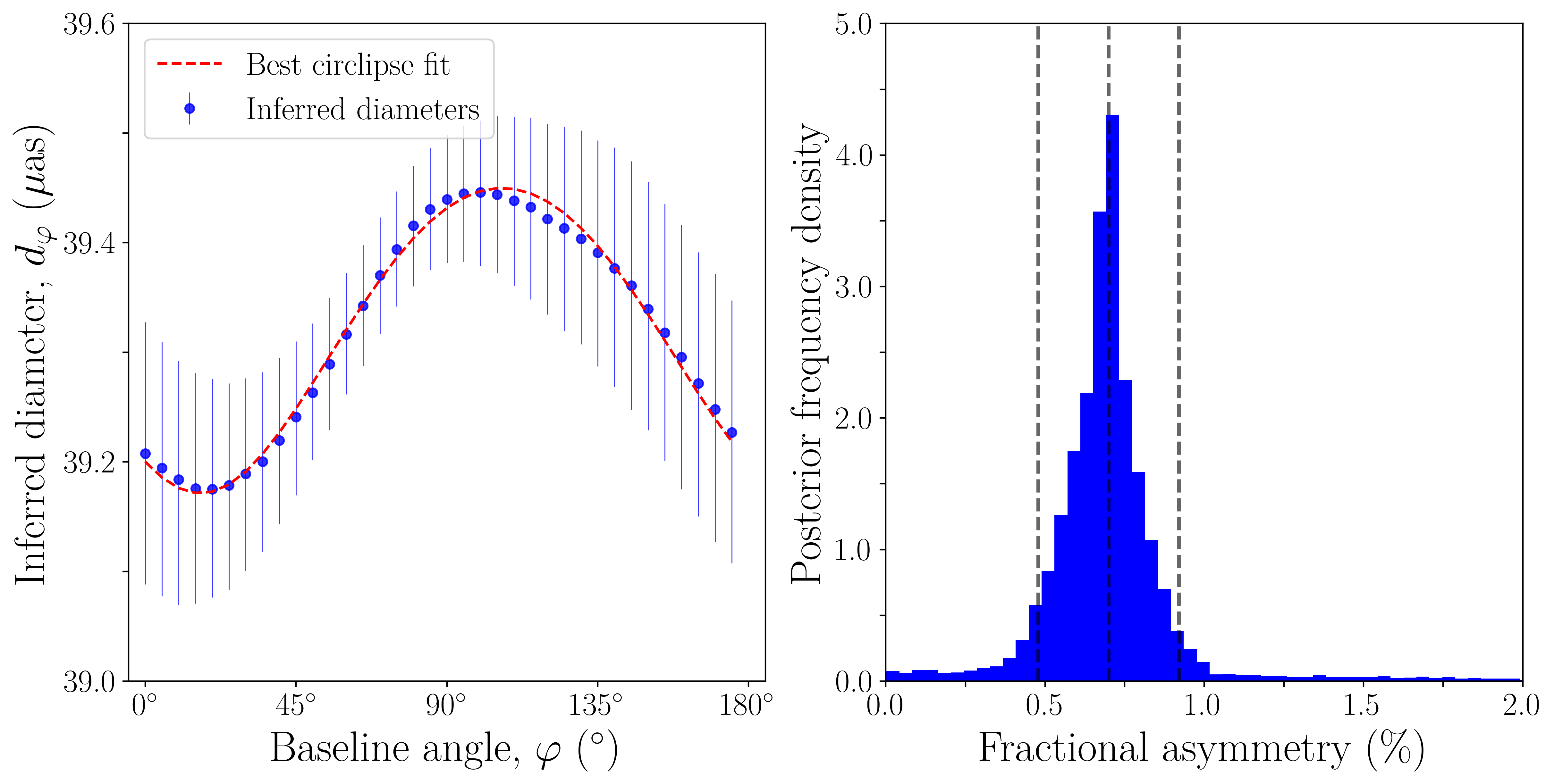}
    \caption{\textbf{Left:} Inferred diameters and their $1\sigma$ error bars (blue), obtained from MCMC-based fits of \texttt{AART}-simulated visibility amplitudes along the $36$ cuts $\varphi=\{0^{\circ}, 5^{\circ},\ldots,175^{\circ}\}$ across the baseline window $[20, 40]\,\mathrm{G}\lambda$ (with artificially-prescribed errors as explained in Sec.~\ref{ssec:SpinInferenceGLM}). In dashed red we plot the best-fit circlipse~\eqref{eq:Circlipse}. The underlying black hole has spin $a_{*}=0.94$, inclination $i_{*}=17^{\circ}$ and $J_{\mathrm{SU}}$ emission profile parameterized by $\mu=r_{-}$, $\sigma=0.5\,M$, and $\gamma=-1.5\,M$. \textbf{Right:} Posterior distribution (blue) of the fractional asymmetry, obtained via pointwise evaluation of Eq.~\ref{eq:FractionalAsymmetry} using the posterior distributions of the circlipse parameters $(R_{0}, R_{1}, R_{2})$. The black dashed lines correspond to the $2.5\%$, $50\%$ and $97.5\%$ percentiles of the posterior distribution, given by $0.48\%$, $0.70\%$, and $0.92\%$, respectively.}
\label{fig:GLMPosteriors}
\end{figure*}

Given the simulated visibility amplitude and errors, we run MCMC simulations to obtain posterior distributions of the visibility amplitude fit parameters (\ref{eq:UniversalVisibility}), using a Gaussian likelihood (\ref{eq:VisibilityLikelihood}) which penalizes points with a larger relative uncertainty. In this case, the posterior distribution of each diameter $d_{\varphi}$ is single-peaked---we plot in the left panel of Fig.~\ref{fig:GLMPosteriors} the inferred diameters $d_\varphi$ and their corresponding $1\sigma$ errors, $\sigma_{\varphi}$, obtained from fits across the baseline window $[20, 40]\,\mathrm{G}\lambda$. The inferred diameters correspond to the $50$th percentile of each posterior distribution, obtained by fitting the distribution to a Gaussian functional form whose width gives the corresponding $1\sigma$ error.

\begin{figure*}
    \centering
    \includegraphics[width=\textwidth]{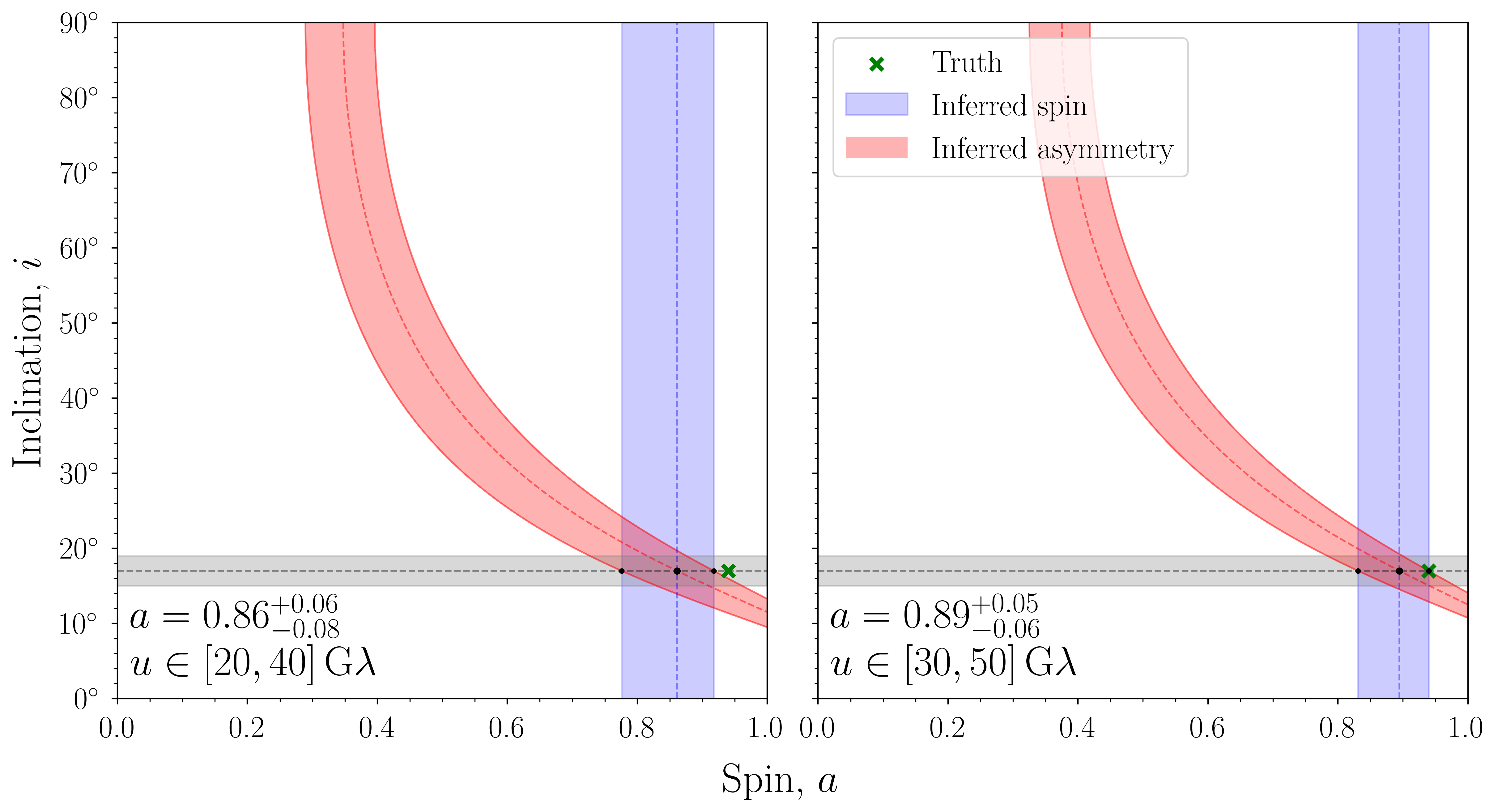}
    \caption{$2\sigma$ inferred asymmetry contour bands (red) with leftmost, central, and rightmost contours given, respectively, by the $2.5\%$, $50\%$ and $97.5\%$ percentiles of the fractional asymmetry posterior distribution, shown in the right panel of Fig.~\ref{fig:GLMPosteriors} for the baseline window $[20, 40]\,\mathrm{G}\lambda$ (left), and obtained similarly for $[30, 50]\,\mathrm{G}\lambda$ (right). We illustrate spin inference when assuming the true underlying inclination $i_{*}=17^{\circ}$ is known exactly: the intersection between the grey dashed horizontal line $i=17^{\circ}$ with each asymmetry contour, shown as black dots, provide the $2\sigma$ prediction for the spin shaded in blue. The grey shaded region corresponds to the inclination prior $i=17^{\circ}\pm2^{\circ}$, in which case the resulting $2\sigma$ spin inference has a lower bound given by that when repeating the above procedure for $i=19^{\circ}$, a central value given by that when assuming $i=17^{\circ}$, and upper bound given by that when assuming $i=15^{\circ}$.}
\label{fig:GLMSpinInference}
\end{figure*}

The pairs $(d_{\varphi}, \sigma_{\varphi})$ are then used as input to another MCMC simulation (with likelihood given by Eq.~\ref{eq:CirclipseLikelihood}) to obtain posterior distributions of the circlipse parameters (\ref{eq:Circlipse}), from which the posterior distribution of the fractional asymmetry is obtained by pointwise evaluation of Eq.~\ref{eq:FractionalAsymmetry}. The resulting best-fit circlipse is plotted in red in the left panel of Fig.~\ref{fig:GLMPosteriors}, while in the right panel we plot the posterior distribution of the fractional asymmetry with vertical dashed lines corresponding to a $2\sigma$ confidence interval, i.e., the $2.5\%$, $50\%$ and $97.5\%$ percentiles, in this case given by asymmetry values $0.48\%$, $0.70\%$, and $0.92\%$, respectively. These values of the shape asymmetry are then used as input to the predictive grid (shown in Fig.~\ref{fig:AsymmetryContours}) to infer spin---we plot in Fig.~\ref{fig:GLMSpinInference} inference bands which span the $2\sigma$ confidence interval of the fractional asymmetry, with the central dashed contour corresponding to the $50$th percentile.

We illustrate in Fig.~\ref{fig:GLMSpinInference} the spin inference procedure when assuming we know the underlying inclination $i_{*}=17^{\circ}$ exactly. The intersection of the $2.5\%$, $50\%$ and $97.5\%$ percentile contours with the horizontal line $i=17^{\circ}$ in the spin-inclination plane give a $2\sigma$ prediction of the black hole spin, which is $a=0.86^{+0.06}_{-0.08}$ from fits across the window $[20, 40]\,\mathrm{G}\lambda$, and $a=0.89^{+0.05}_{-0.06}$ for the same analysis repeated across $[30, 50]\,\mathrm{G}\lambda$---these $2\sigma$ ranges are shaded in blue in Fig.~\ref{fig:GLMSpinInference}. To account for the uncertainty in the inclination, we repeat the same process with assumed inclinations $i=15^{\circ}$ and $i=19^{\circ}$, obtaining for each a $2\sigma$ interval for the inferred spin. The inferred spin with an inclination prior $i=17^{\circ}\pm2^{\circ}$ then has a central value given by that when assuming $i=17^{\circ}$, while the lower and upper bound on the $2\sigma$ confidence interval are given by those when assuming $i=19^{\circ}$ and $i=15^{\circ}$, respectively. We list in Tab.~\ref{tbl:GLM_SpinInferenceStatistics} the inferred spin when assuming the inclination prior is exact (first column) and when accounting for its $2\sigma$ confidence interval (second column). Results are not presented for the baseline window $[80, 100]\,\mathrm{G}\lambda$ since, across this window, the visibility amplitude is in a transition between $n=1$ and $n=2$ for certain baseline angles (e.g., see Fig.~3 of Ref.~\cite{CardenasAvendano_2024}), causing the circlipse inference (and hence the spin inference) to fail in this case.

\begin{table}
\renewcommand{\arraystretch}{2.5}
\begin{tabular}{ccc}
\cline{2-3}
\multicolumn{1}{c|}{} & \multicolumn{1}{c|}{$i=17^{\circ}$} & \multicolumn{1}{c|}{$i=17^{\circ}\pm 2^{\circ}$}\tabularnewline
\hline
\multicolumn{1}{|c|}{\shortstack{Baseline\\Window}} & \multicolumn{1}{c|}{\shortstack{Inferred\\Spin}} & \multicolumn{1}{c|}{\shortstack{Inferred\\Spin}}\tabularnewline
\hline
\hline
$[20, 40]\,\mathrm{G}\lambda$ & $0.86^{+0.06}_{-0.08}$ & $0.86^{+0.1}_{-0.13}$ \tabularnewline
$[30, 50]\,\mathrm{G}\lambda$ & $0.89^{+0.05}_{-0.06}$ & $0.89^{+0.1}_{-0.11}$ \tabularnewline
$[280, 300]\,\mathrm{G}\lambda$ & $0.94^{+0.03}_{-0.04}$ & $0.94^{+0.06}_{-0.08}$ \tabularnewline
\hline
\hline
\end{tabular}\caption{Spin inference results using the MCMC-based inference method described in Sec.~\ref{sssec:MCMC} on simulated visibility amplitudes with uniform pointwise measurement uncertainty given by $33\%$ of the signal power for a given radon cut $\varphi$ across a given baseline window. The underlying black hole has spin $a_{*}=0.94$, inclination $i_{*}=17^{\circ}$ and $J_{\mathrm{SU}}$ emission profile parameterized by $\mu=r_{-}$, $\sigma=0.5\,M$, and $\gamma=-1.5\,M$. The first column corresponds to spin inference when assuming the inclination is known exactly, i.e., $i = i_{*}$, while the second column assumes the confidence interval $i=i_{*}\pm2^{\circ}$.}
\label{tbl:GLM_SpinInferenceStatistics}
\end{table}

The central values of the spin inferred across the baseline windows $[20, 40]\,\mathrm{G}\lambda$, $[30, 50]\,\mathrm{G}\lambda$, and $[280, 300]\,\mathrm{G}\lambda$ are $9\%$, $5\%$, and $0\%$, respectively, lower than the true spin value (i.e., there is a systematic offset), while the total width of the $2\sigma$ confidence intervals are $16\%$, $12\%$ and $7\%$, respectively, of their central values. Only at the $3\sigma$ level does the $[20,40]\,\mathrm{G}\lambda$  inferred spin confidence interval contain the true value---this interval is given by $0.86^{+0.08}_{-0.15}$, which has a total width which is $27\%$ of its central value. We see from the final column in Tab.~\ref{tbl:GLM_SpinInferenceStatistics} that the effect of accounting for the $2\sigma$ error in the inclination is to increase the width of the $2\sigma$ confidence interval for the inferred spin by a factor of approximately two.

\subsection{Black Hole Spin Inference from Synthetic Observations}
\label{ssec:SpinInferenceBHEX}

Thus far we have considered simulations based on time-averaged images. We now consider black hole movies generated by \texttt{AART} in slow-light mode. The time and spatial variability in the model outlined in Sec.~\ref{Sec:TheoreticalFramework} are driven by an inhomogeneous, anisotropic, and time-dependent Gaussian random field (GRF), which is modeled using the \texttt{inoisy} code~\cite{Lee:2020pvs}. The spatio-temporal correlations of this field follow a Keplerian-like scaling, meaning that the correlation structure of the variability is proportional to the spatial structure. For further details, we refer the reader to Ref.~\cite{CardenasAvendano_2022}.

Modeling photon ring fluctuations as a Gaussian random field with a Matérn covariance cannot capture the rich dynamics revealed in fluid-based simulations of black-hole accretion flows, e.g., intermittent, burst-like events driven by magnetic reconnection and eddies~\cite{Ripperda:2021zpn}. Consequently, using a Gaussian-based fluctuation model can miscalculate variability and introduce subtle systematic shifts in fitted parameters or image morphology arising from non-Gaussian brightness excursions~\cite{Chang:2025hrk}. Fortunately, over longer timescales, the time-averaged image structure, the regime in which we will apply this spin inference scheme, remains relatively insensitive to the specific turbulence and variability models. However, the degree of averaging and the campaign's details will depend on the nature of the variability and are worth studying in future works.

Following the notation of Ref.~\cite{CardenasAvendano_2022}, we assume the particles in the disk follow pure geodesic circular Keplerian orbits and the following  correlation scales: $\lambda_0=2\pi/\Omega_{\rm K}$, $\lambda_1=5r_{\rm s}\,,\lambda_2=0.1r_{\rm s}$, where $r_{\rm s}$ denotes the source radius and $\Omega_{\rm K}$ the Keplerian frequency. The anisotropy direction (the opening angle of the spiral features of the emission) is $\theta_\angle=20^\circ$. The underlying \texttt{inoisy} simulations were run on a regular Cartesian grid $(t_{\rm s},x_{\rm s},y_{\rm s})$ of size
$2048\times1024\times1024$. Specifically, for each of the spatial coordinates $(x_{\rm s},y_{\rm s})$ in the equatorial disk, we uniformly placed
$1024$ pixels within the range $[-30\,M,30\,M]$, resulting in a resolution of about $0.06\,M$, whereas for the time coordinate we placed $2048$ grid points uniformly distributed within the range $[0\,M,5000\,M]$, resulting in a cadence of $2.44\,M$.

Once a realization of the GRF, $\mathcal{F}(x,y,t)$, is generated, black hole movies are produced by multiplying the source intensity, $I_{\rm s} = J_{\rm SU}$, by this inhomogeneous, anisotropic, time-dependent field, as follows~\cite{CardenasAvendano_2022}:
\begin{equation}
    I_{\rm s}(x,y,t) \Rightarrow I_{\rm s}(r) \times \exp\left[\sigma_{\rm scale} \times \mathcal{F}(x,y,t) - \sigma_{\rm scale}^{1/2}\right],
\end{equation}
where $\sigma_{\rm scale}$ is a fluctuation scale factor that controls the variability of the fluctuations. When $\sigma_{\rm scale} = 0$, the model reduces to the time-averaged version described in Sec.~\ref{Sec:TheoreticalFramework}.

We study two simulations, both with $a_{*}=0.94$ and $i_{*}=17^{\circ}$: Simulation A, with parameters $(\mu=r_{-}, \vartheta=0.5\,M, \gamma=-1.5\,M)$ and $\sigma_{\rm scale}=0.4$; and Simulation B, with parameters $(\mu=1.5\,M, \vartheta=1.0\,M, \gamma=0)$ and $\sigma_{\rm scale}=0.3$. Simulation A corresponds to the emissivity profile studied in Sec.~\ref{ssec:SpinInferenceGLM}, while Simulation B uses a different profile that reaches the ``universal regime'' more quickly and exhibits less variability. From these two simulations we mock an observation using \texttt{ngEHTsim}~\cite{2024ApJ...968...69P}, a VLBI synthetic data pipeline built around \texttt{eht-imaging}~\cite{Chael_2016} that models realistic weather and instrument corruptions to interferometric data. In each case, we rescale each movie so that the average total flux at each frequency matches the benchmark values in \citet{Johnson:2024ttr}.

We simulate a BHEX-like campaign on M87* using three months of observations every three days, beginning January 1st, 2031. We assume that the satellite component is in a circular polar orbit with a 12 hour period, with right ascension of the ascending node chosen to put the orbital plane face-on to M87* (approximately $277^\circ$). These observations assume gain amplitudes are estimable, but assumes no gain phase information is recovered. The resulting error is thermal noise on the complex visibility observed on a single baseline, which includes contributions from both instrumental and atmospheric contributions at each station. We simulate observations with \texttt{ngEHTsim}~\cite{Pesce_2024} using Frequency Phase Transfer (FPT) from a low band tuned to $86$ GHz (hereafter, the ``low band''), and a high band with two sidebands, centered at $251.5$ GHz and $267.5$ GHz (hereafter, the ``high band''). This setup follows the projected receiver architecture outlined in Ref.~\cite{Tong_2024}. 

\begin{figure*}
    \centering
    \includegraphics[width=\textwidth]{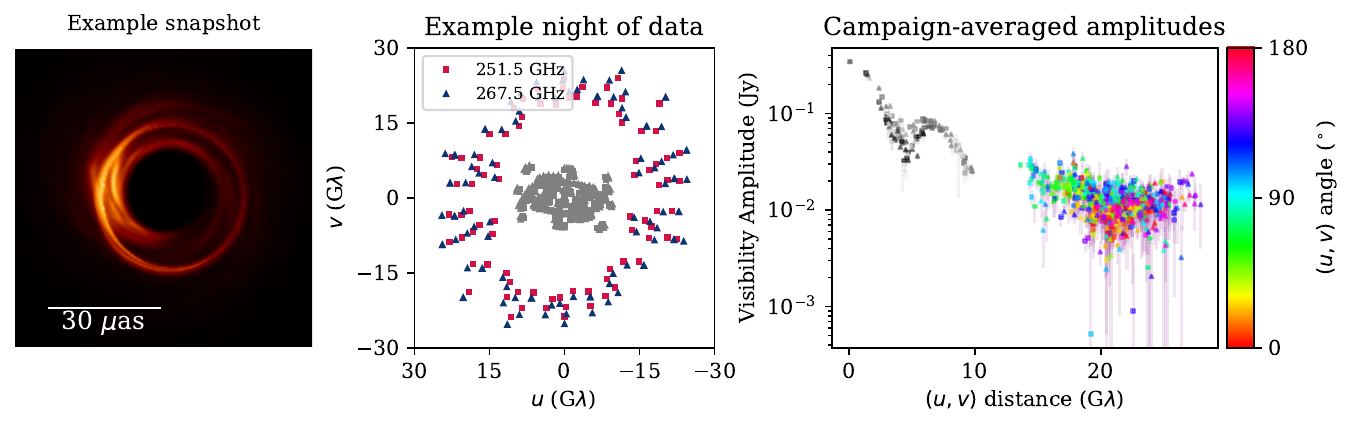}
    \caption{Example synthetic data from the campaign observation pipeline. Left: example image from the \texttt{AART} movie (Simulation B). Middle: example single-night $(u,v)$ coverage observing the image on the left. Right: corresponding campaign-averaged visibility amplitudes, constructed from $30$ nights of observations every three days starting January 1, 2030. Gray points show ground data, and error bars on the right indicate errors on the mean amplitude. For the analysis presented in Sec.~\ref{ssec:SpinInferenceBHEX}, we have only used the space-based data.}
    \label{fig:synthetic_data_example}
\end{figure*}

These simulations assume co-observation from VLBI observatories on Earth, namely the Green Bank Telescope (low band only), Haystack Observatory (low band only), Yebes Observatory (low band only), the IRAM 30m Telescope (low and high bands), the Submillimeter Telescope (high band only), the James Clark Maxwell Telescope (low and high band), the Submillimeter Array (high band only), the Northern Extended Millimetre Array (low and high band), the Greenland Telescope (high band only), the Large Millimeter Telescope (low and high band), the Korean VLBI Network's Pyeongchang and Yonsei dishes (low and high band), and the African Millimeter Telescope (low and high band) in its planned Namibia location. An example of a snapshot from Simulation B and the resulting interferometric data are shown in Fig.~\ref{fig:synthetic_data_example}. 

It is important to emphasize that the campaign parameters adopted here serve primarily to demonstrate an end-to-end inference pipeline rather than to prescribe an optimal observing strategy. Identifying the ideal combination of array geometry, cadence, bandwidth, or, for instance, campaign duration---while balancing gains in sensitivity against the risk of long-term source evolution---requires a dedicated optimization study that we leave to future work. It also bears emphasizing that efforts to simulate realistic BHEX data are an ongoing work; the tools and data realizations we work with here are preliminary. For example, \texttt{ngEHTsim} does not yet include closure triangle phase steering across bands, causing the number of successful integrations on high band-only baselines to be drastically underestimated when high sensitivity low band sites are present. This functionality is currently under development.

In order to estimate average properties of the sky morphology, we average data across the campaign using a greedy scheme that seeks clusters of $(u,v)$ points. In detail, we first choose a evaluation radius $\rho_{\rm eval}\equiv0.5$G$\lambda$ over which we wish to gather amplitudes for averaging. We then convolve the measured $(u,v)$ points in the campaign with a disk of radius $\rho_{\rm eval}$ and rasterize the convolution onto a grid of $0.1$ G$\lambda$ resolution. The ``greedy'' averaging corresponds to finding peaks in this density map, averaging all observations within $\rho_{\rm eval}$ of each peak, removing these points from the density map, and continuing until only single measurements remain. During averaging, amplitudes are weighted by the inverse of the combined sample and thermal variance after cutting integrations with an estimated thermal noise greater than $10$ mJy; in this way, long integrations dominated by bright skies in poor weather are removed, while intrinsic variation smooths contributions from observations with different signal-to-noise ratio. Lastly, we separate the resulting visibilities into angular bins of size $10^\circ$ in the $(u,v)$ plane, giving a total of $18$ Fourier angles (down from $36$ as obtained from the \texttt{AART} simulations in the previous sections). For the following analysis, we will only consider data from the high band. 

\begin{figure*}
    \centering
    \includegraphics[width=\textwidth]{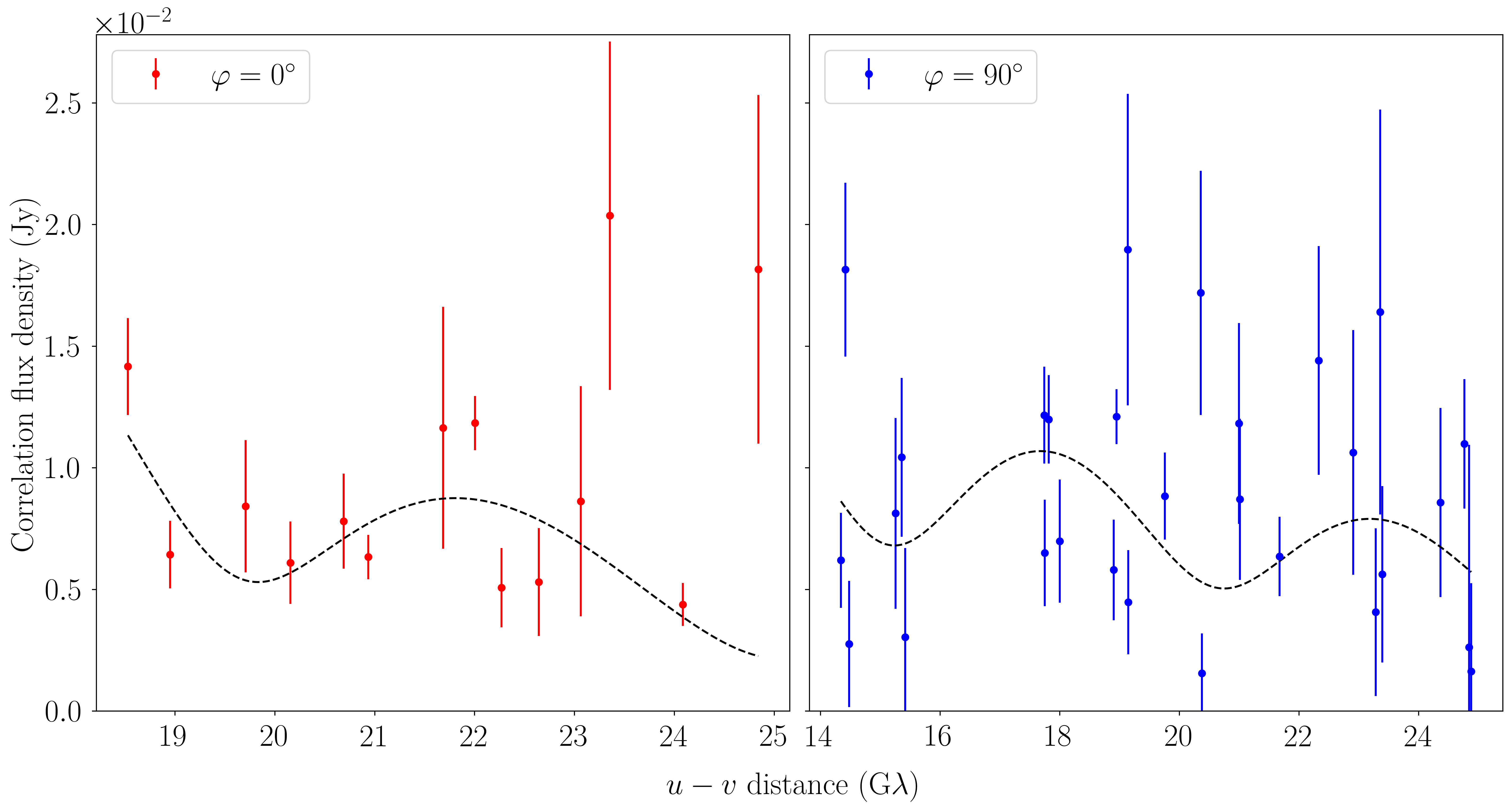}
    \caption{Simulation A binned visibility amplitude along cuts $\varphi=0^{\circ}$ (left) and $\varphi=90^{\circ}$ (right), their associated errors, and their MCMC-based fits to the functional form~\eqref{eq:VisibilityFit} (black dashed).}
\label{fig:BHEXVisampInference}
\end{figure*}

We plot in Fig.~\ref{fig:BHEXVisampInference} the binned visibility amplitude data along cuts $\varphi=0^{\circ}$ (left) and $\varphi=90^{\circ}$ (right) for Simulation A. In addition to the more comprehensive sources of noise and uncertainty and the reduction in the number of Fourier angles from $36$ to $18$, this data differs from the time-averaged simulations in the previous sections in that the visibility amplitudes along the various cuts no longer all span the same baseline window.

\begin{figure*}
    \centering
    \includegraphics[width=\textwidth]{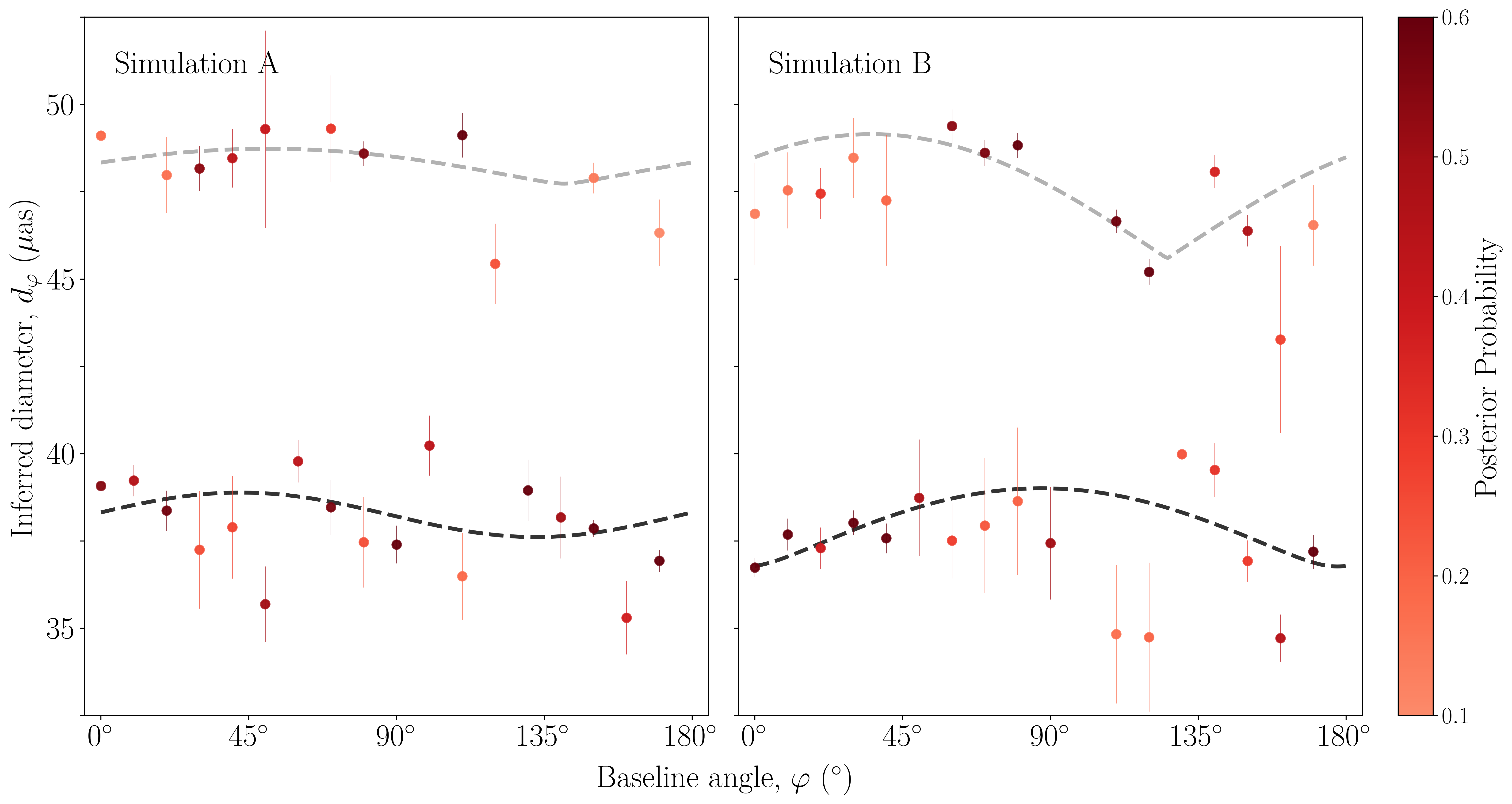}
    \caption{Candidate diameters (scatter points) for Simulation A (left) and Simulation B (right), obtained by fitting all peaks in the posterior distribution for $d_\varphi$ along each cut $\varphi\in\{0^{\circ}, 10^{\circ},\ldots,170^{\circ}\}$ to a Gaussian functional form, whose width determines the $1\sigma$ error bars for each diameter, and whose total area gives the posterior probability associated with a given diameter as indicated by the colorbar. We plot in black the best MCMC-based fits to the circlipse functional form~\eqref{eq:Circlipse}, and in the darker shade the most probable of the candidate circlipses for each simulation.}
\label{fig:BHEXCirclipses}
\end{figure*}

Having averaged and binned the visibility data, the spin inference approach is broadly the same as in Sec.~\ref{ssec:SpinInferenceGLM}. We first infer diameters $d_\varphi$ and their $1\sigma$ error from their posterior distributions---the best visibility amplitude fits for $\varphi=0^{\circ}$ and $\varphi=90^\circ$ are plotted as black dashed curves in Fig.~\ref{fig:BHEXVisampInference}. As was the case for the baseline window $[280, 300]\,\mathrm{G}\lambda$ in Sec.~\ref{ssec:SpinInferenceGLM}, we obtain multi-peaked posterior distributions of the diameters $d_{\varphi}$, yielding two candidate circlipses for each simulation as shown in Fig.~\ref{fig:BHEXCirclipses}---the color bar therein corresponds to the posterior probability of each diameter. 

We see in Fig.~\ref{fig:BHEXCirclipses} that for the synthetic observations, the multi-peaked posterior distributions of the diameters $d_{\varphi}$ no longer all contain the same number of peaks; across certain angles $\varphi$, some candidate circlipses do not have an inferred diameter. To determine the most probable of the two candidate circlipses for a given simulation, we first identify the subset of angles $\varphi$ for which each circlipse has a constituent diameter (i.e., the angles for which there are exactly two candidate diameters). For each circlipse, we compute the product of the diameter posterior probabilities over these angles, and identify the circlipse with the largest product as the most probable. We then use the distribution $(d_{\varphi}, \sigma_{\varphi})$ of this circlipse as input to the MCMC simulation for the circlipse fit parameters. We plot as black dashed curves in Fig.~\ref{fig:BHEXCirclipses} the best-fit curves to each candidate circlipse, with the darkest curve being the fit to the most probable circlipse.

\begin{figure}
    \centering
    \includegraphics[width=\columnwidth]{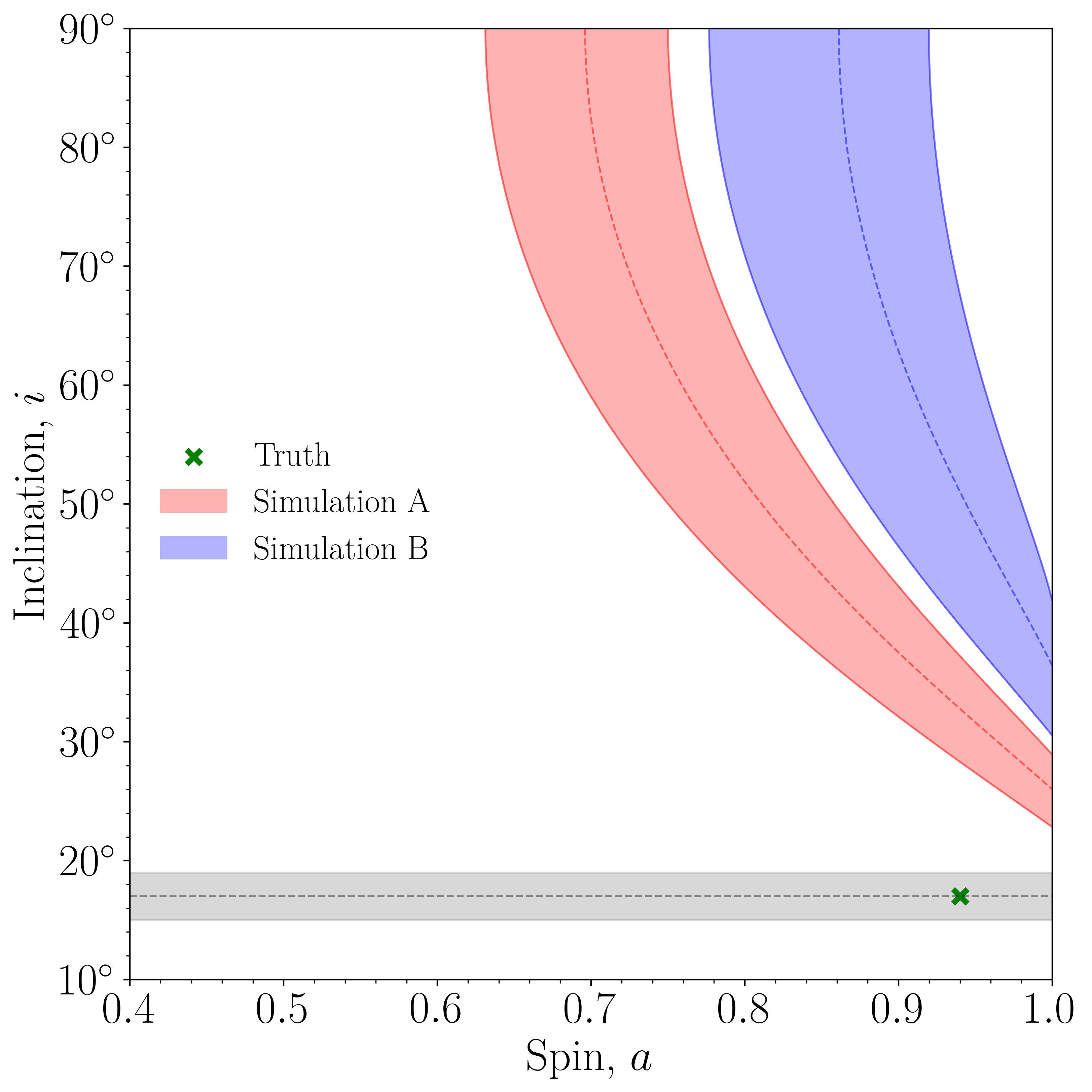}
    \caption{$2\sigma$ inferred asymmetry bands for Simulation A (red) and Simulation B (blue). The inner, central (dashed), and outer contours for each band correspond to the $2.5\%$, $50\%$ and $97.5\%$ percentiles of the fractional shape asymmetry posterior distribution, given, respectively, by $2.6\%$, $3.3\%$, and $4.0\%$ for Simulation A, and by $4.4\%$, $5.9\%$, and $7.5\%$ for Simulation B. The grey shaded region denotes the inclination prior $i=17^{\circ}\pm2^{\circ}$.}
\label{fig:BHEXParamInference}
\end{figure}

From the posterior distribution of the best circlipse fit parameters for each simulation, we obtain a single-peaked fractional asymmetry posterior distribution. We plot in Fig.~\ref{fig:BHEXParamInference} asymmetry bands corresponding to a $2\sigma$ confidence interval for the inferred asymmetry for each simulation. In both cases, the bands do not intersect with any horizontal lines of fixed inclination for $i\leq19^{\circ}$, meaning that our spin inference method fails to give a prediction for the spin with the inclination prior $i=17^{\circ}\pm2^{\circ}$.

For the time-averaged visibility amplitudes studied in Sec.~\ref{ssec:SpinInferenceGLM}, we inferred a $2\sigma$ confidence interval $f_{A}\in[0.48\%,\,0.92\%]$. In contrast, the ring shape inference scheme applied to the mock simulation A yielded an increased asymmetry estimate $f_{A}\in[2.6\%,\,4.0\%]$ at the $2\sigma$ level. This jump shifts the red inference bands in Fig.~\ref{fig:GLMSpinInference} upward to those shown in Fig.~\ref{fig:BHEXParamInference}, where our critical-curve-based inference scheme finds such a high shape asymmetry to be incompatible with the prior inclination $i=17^{\circ}\pm2^{\circ}$. Nevertheless, we can still place lower bounds on the spin from these synthetic observations. Specifically, the lowest spin consistent with each inferred asymmetry band in Fig.~\ref{fig:BHEXParamInference} is given by the $x$-coordinate where the left edge of the band intersects the horizontal line $i=90^\circ$. For Simulations A and B, this yields lower bounds on the spin of $0.64$ and $0.68$, respectively, at the $2\sigma$ level and $0.59$ and $0.72$ at the $3\sigma$ level. 

With the chosen parameters for \texttt{ngEHTsim} and the inference methods we used, we were unable to infer a distribution of the spin parameter from Simulation A nor Simulation B, as was done in previous examples. There may be multiple reasons for this, and, in this work, we did not specifically tune the parameters to achieve a stronger spin constraint using our method. Instead, the two emission models, along with the selected \texttt{ngEHTsim} parameters, primarily serve as a proof-of-concept application of the particular ring shape and parameter inference schemes employed in this paper. While the observation parameters are based on educated assumptions, their optimal values remain to be determined in future studies. Consequently, the results presented here should be viewed as illustrative examples highlighting the methodology, rather than a demonstration of the inference capabilities of a potential BHEX-like mission. 

We emphasize that a less variable model could have been chosen since we do not yet know whether future-targeted sources will exhibit the level of variability assumed here. Similarly, our chosen number of observation nights, frequencies, and Earth-based nodes may not be fully representative of the eventual telescope configuration. Moreover, there are caveats associated with quoting these lower bounds. A sufficiently noisy dataset might inflate the inferred shape asymmetry, pushing the inference bands toward larger spin values even when the true spin is lower. In a less extreme scenario---as noted in Sec.~\ref{sec:AstroStudy}---the asymmetry for lower intrinsic spins can still be overestimated at baselines $\sim 300\,\mathrm{G}\lambda$, causing the true spin to lie outside the stated lower bound. Therefore, in presenting these lower bounds, we implicitly assume that when using our method: (a) spin inference is carried out in the low-inclination, rapidly-rotating regime where the method performs best (Sec.~\ref{sec:AstroStudy}); and (b) although the fractional asymmetry constraints in this section are inaccurate and do not include those obtained in the noiseless case, they nevertheless remain informative insofar as the visibility amplitude and circlipse fits are not rendered meaningless by excessive noise. 

\section{Discussion and Future Outlook}
\label{Sec:Discussion}

We have presented a geometric, astrophysics-agnostic approach to inferring the spin of rapidly-rotating, low-inclination black holes from the interferometric shape of the first photon ring. We applied this method to simulations of the interferometric signature of black holes with viewing inclination $i_{*}=17^{\circ}$, spins $a_{*}\in\{0.5, 0.94\}$, and for $140$ different emission models for the equatorial emission. 

For noise-free, time-averaged images in the high-spin ($a_{*}=0.94$) case, we found that when varying the underlying model of emission, the inference method applied at very long baselines ($\sim300 \,\mathrm{G}\lambda$)---where the second photon ring typically dominates---yields a distribution of spin values centered on the true value with a $2\sigma$-width of $5\%$ of the spin value. At shorter baselines typically dominated by the first photon ring, the central value of the inferred spin distribution was $5\%$ less than the true value and the distribution was approximately $3$ times as wide as in the (very) long-baseline case. The shift towards an underestimation of the spin at shorter baselines is due to the inferred shape of the first photon ring being less asymmetric than the critical curve shape for the majority of profiles considered here. 

When applying our method to synthetic observations for a BHEX-like mission (again for the $a_{*}=0.94$ case), our shape inference scheme was unable to provide a precision estimate of the photon ring's fractional shape asymmetry, instead estimating it to be $\sim3$ times as large as in the noiseless case, which caused the spin inference method to break down and only provide lower bounds on the spin parameter. 

For the profiles here considered, the method also breaks down in the case $a_{*}=0.5$ in the absence of noise. At baselines where the second photon ring typically dominates, the inference is less accurate than the high-spin case and \emph{much} less precise---the average inferred spin in this case was $14\%$ larger than the true value with a $2\sigma$ error interval of total width given by $76\%$ of the true spin value. The method is similarly imprecise at shorter baselines but much less accurate (i.e., at the shortest baselines considered, the average inferred spin was $\sim{50}\%$ larger than the true value). The failure of our spin-inference method in the slowly-rotating case is due to the increased degeneracy of the critical curve shape at such values of the spin parameter and less accurate ring shape inferences, which also become more dependent on the choice of emission profile. 

In order to be conservative in this study, we considered a wide range of emission profiles without considering \emph{any} constraints provided by existing interferometric observations. For the majority of profiles here considered, including that in Sec.~\ref{ssec:SpinInferenceGLM}, which is broadly consistent with short-baseline EHT measurements, the inferred photon ring shape asymmetry systematically underestimates that of the critical curve. If this is a shared feature of more realistic models of emission, one might be able to augment our approach with a scheme to address this systematic bias. For example, one might be able to study photon ring shapes over a range of realistic emission profiles and introduce a correction factor to account for the underestimated asymmetry relative to that of the critical curve. Further, as mentioned in the introduction, here we have used in isolation the fractional shape asymmetry of the photon ring for spin inference while future space-based observations will measure additional observables which, in combination, would provide better constraints on the black hole spin. 

While our analysis highlights challenges---particularly in noisy conditions and for low-spin black holes---it represents an important first step toward achieving precise, model-agnostic spin measurements. Our method uses only a minimal subset of the available observables instead of exploiting the full richness of the photon ring data that future space-based missions will provide. Consequently, this approach does not reflect the complete spin inference methodology anticipated for future missions. 
In particular, we anticipate the explicit, joint modeling of the $n=0$ and $n=1$ image morphology, rather than the critical curve itself, to be crucial for reducing systematics in estimates of spacetime parameters from data in the $[20, 40]$~G$\lambda$ regime. Furthermore, incorporation of multi-frequency-band observations, calibration with realistic emission model libraries, development of improved shape descriptors, and leveraging polarimetric information from the $n=0$ and $n=1$ images are all expected to significantly enhance the robustness and accuracy of spin measurements.

\acknowledgments

We thank Alexandru Lupsasca, Andrew Chael, Dorothy Cutler, George Wong, and the BHEX Photon Ring Science Group for many helpful discussions. A.\;C.-A. is supported by LANL Laboratory Directed Research and Development, grant 20240748PRD1, as well as by the Center for Nonlinear Studies. LANL is operated by Triad National Security, LLC, for the National Nuclear Security Administration of the U.S. DOE (Contract No. 89233218CNA000001). This work is authorized for unlimited release under LA-UR-25-21691. D.C.M.P. acknowledges financial support from the National Science Foundation (AST-2307887). This work was supported by the Black Hole Initiative, which is funded by grants from the John Templeton Foundation (Grant 62286) and the Gordon and Betty Moore Foundation (Grant GBMF-8273) - although the opinions expressed in this work are those of the author(s) and do not necessarily reflect the views of these Foundations. The simulations presented in this article were performed on computational resources managed and supported by Princeton Research Computing, a consortium of groups that includes the Princeton Institute for Computational Science and Engineering (PICSciE), and the Office of Information Technology's High Performance Computing Center and Visualization Laboratory at Princeton University.

\appendix

\section{Parameter Inference with a Black Hole Mass Prior}
\label{App:SpinInclinationInference}
In this appendix, we describe an extension of the spin inference method described in this paper to inferring both black hole spin and inclination given a strong prior on the black hole mass. We work in units of the black hole mass $M$, thereby assuming we know its value exactly.

\begin{figure}
    \centering
    \includegraphics[width=\columnwidth]{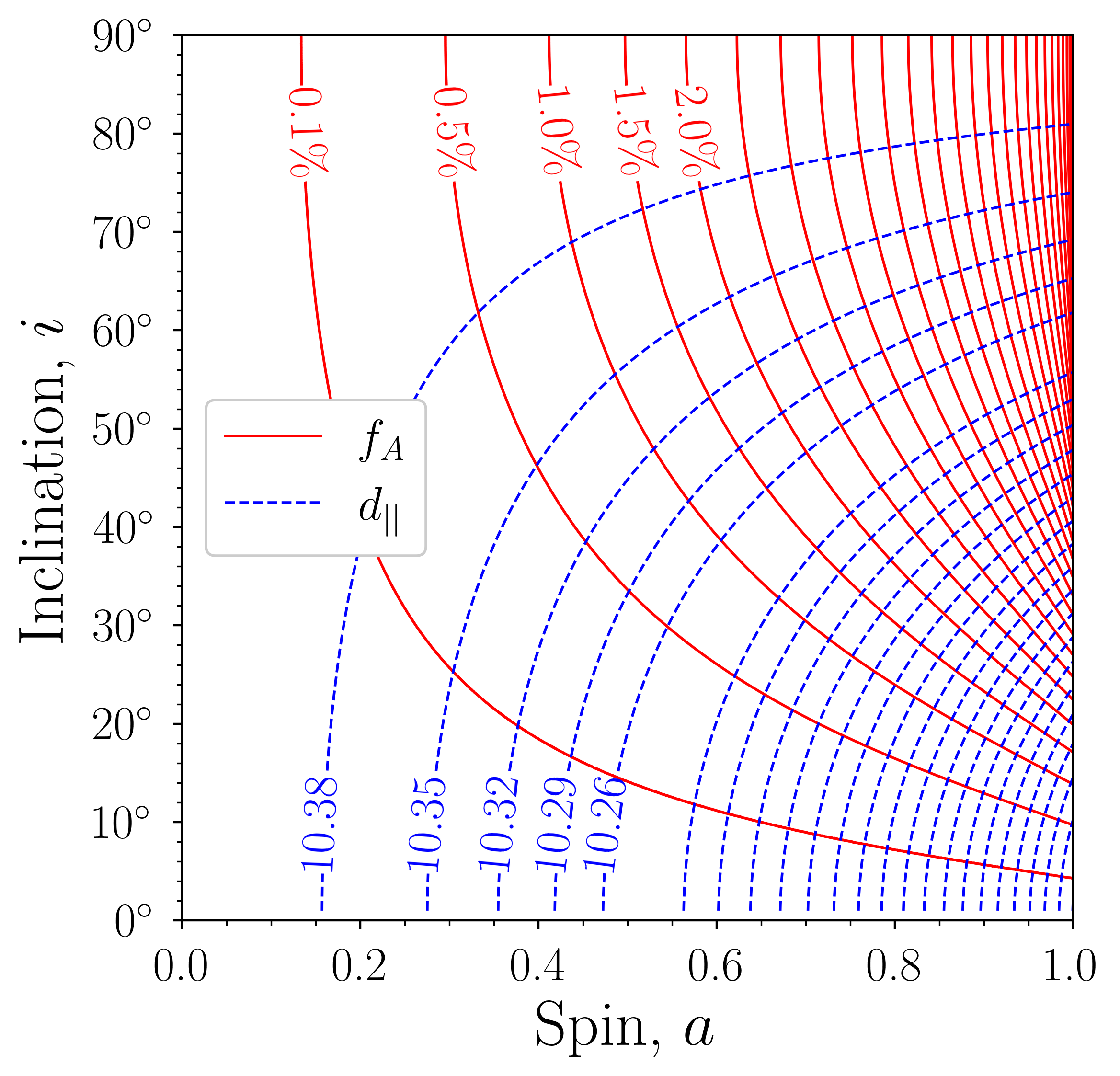}
    \caption{Fractional shape asymmetry (red) and parallel diameter $d_{||}$ (blue, in units of $M$) contours interpolated from critical curves with spin-inclination pairs $(a_{*},i_{*})$, where $a_{*}$ takes on $2000$ equally-spaced values in the range $[0.00001, 0.999999]$, and $i_{*}$ takes on $2000$ values in the range $[1^{\circ}, 90^{\circ}]$. (See also Fig.~7 of Ref.~\cite{Johnson_2020}.) The leftmost asymmetry contour corresponds to a critical curve asymmetry of $0.1\%$ while that of the subsequent contours follows the arithmetic sequence $0.5\%$, $1.0\%,\ldots,13.0\%$. The parallel diameter contours also follow an arithmetic sequence with $d_{||}=10.38M, 10.35M,\ldots,9.66M$.}
\label{fig:DiameterAsymmetryContours}
\end{figure}

Following the discussion presented in Sec.~\ref{ssec:ShapeOfCritCurve}, we plot in Fig.~\ref{fig:DiameterAsymmetryContours} contours of both the fractional shape asymmetry, $f_{A}$, and the parallel diameter, $d_{||}$, of the critical curve in the spin-inclination plane, where, importantly, the diameters are in units of $M$. 

Given some shape $(d_{||}, f_{A})$ of the critical curve, a diameter and an asymmetry contour is defined in the spin-inclination plane (e.g., Fig.~\ref{fig:DiameterAsymmetryContours}). The $x$- and $y$-value of the intersection between these two contours gives (within numerical error) the exact spin and inclination of the underlying black hole whose critical curve on the observer screen has a shape consistent with that given as input.

Parameter inference proceeds by inferring a shape $(d_{||}, f_{A})$ of the photon ring from the total interferometric signal of the corresponding time-averaged black hole image, converting $d_{||}$ from units of microarcseconds to units of $M$, and using this inferred shape as input to the grid depicted in Fig.~\ref{fig:DiameterAsymmetryContours}, which again constitutes an approximation of the critical curve shape with that of the inferred ring shape. The intersection of the resulting diameter and asymmetry contours in the spin-inclination gives a prediction of the black hole spin and inclination.

\begin{figure*}
    \centering
    \includegraphics[width=\textwidth]{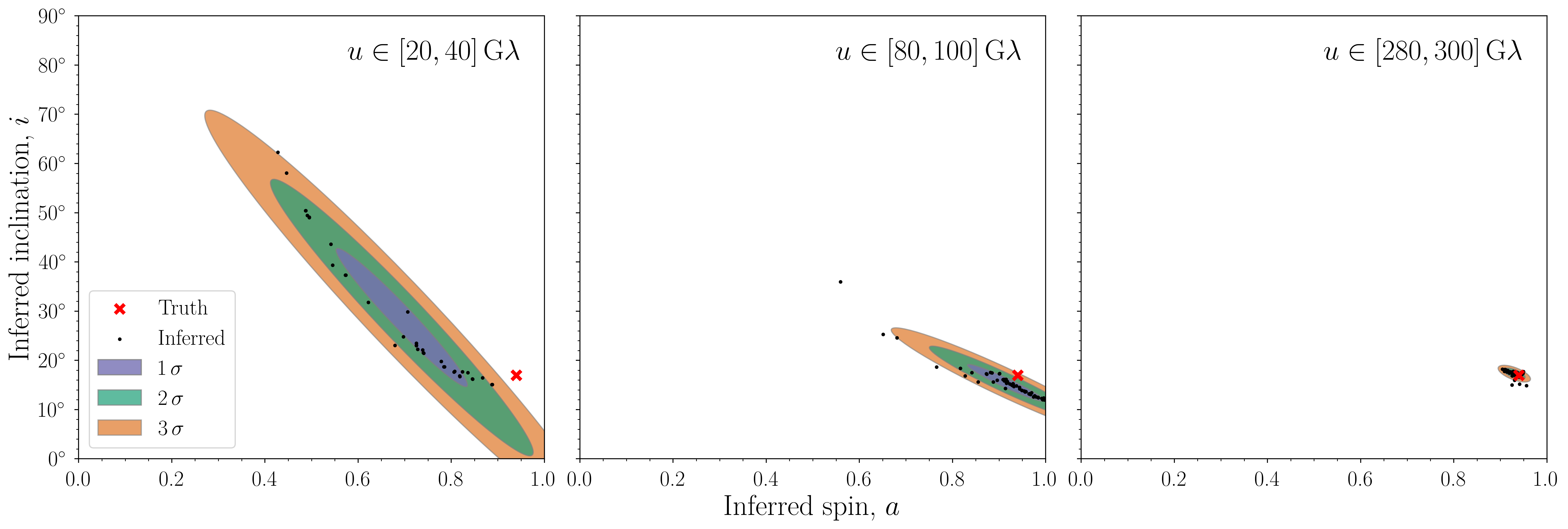}
\caption{Distributions of the spin-inclination pairs inferred across the baseline windows $[20, 40]\,\mathrm{G}\lambda$ (leftmost), $[80, 100]\,\mathrm{G}\lambda$ (center), and $[280, 300]\,\mathrm{G}\lambda$ (rightmost) from noise-free, time-averaged images of black holes with spin $a_{*}=0.94$, inclination $i_{*}=17^{\circ}$, some assumed mass $M$ and $140$ different equatorial emission models as specified in Tab.~\ref{tbl:Parameters}. The percentage of profiles for which a successful spin-inference was made (from left to right) is $26\%$, $42\%$, and $69\%$, respectively. A single inference of spin and inclination is obtained by inferring a diameter-asymmetry pair $(d_{||}, f_{A})$ from the interferometric signal using the least-squares method described in Sec.~\ref{sssec:LeastSquares} and finding the intersection of their associated contours in the spin-inclination plane. We plot the $1\sigma$ (purple), $2\sigma$ (green), and $3\sigma$ (orange) confidence ellipses for each distribution.}
\label{fig:SpinInclinationDistribution}
\end{figure*}

In line with the analysis presented in Sec.~\ref{sec:AstroStudy}, we apply this inference scheme to a noise-free time-averaged imaged of a black hole with underlying spin $a_{*}=0.94$ and inclination $i_{*}=17^{\circ}$ for $140$ astrophysical models of emission specified in Tab.~\ref{tbl:Parameters}. We plot the resulting ($2$D) distributions of inferred spin-inclination pairs across the baseline windows $[20, 40]\,\mathrm{G}\lambda$, $[80, 100]\,\mathrm{G}\lambda$, and $[280, 300]\,\mathrm{G}\lambda$ in Fig.~\ref{fig:SpinInclinationDistribution} and list summary statistics in Tab.~\ref{tbl:AstroSpinIncInferenceStatistics}.

\begin{table}
\resizebox{\columnwidth}{!}{
\renewcommand{\arraystretch}{2.5}
\begin{tabular}{cccc}
\cline{2-4} \cline{3-4} \cline{4-4} 
\hline
\multicolumn{1}{c|}{\shortstack{Baseline\\Window}} & \multicolumn{1}{c|}{\shortstack{Inferred\\Spin}} & \multicolumn{1}{c|}{\shortstack{Inferred\\Inclination}} & \multicolumn{1}{c}{\shortstack{Successful\\Profiles}}\tabularnewline
\hline
\hline
$[20, 40]\,\mathrm{G}\lambda$ & $0.71^{+0.16}_{-0.26}$ & $27.7^{+31.0}_{-12.0}$ & $26\%$ \tabularnewline
$[80, 100]\,\mathrm{G}\lambda$ & $0.92^{+0.07}_{-0.27}$ & $15.2^{+10.2}_{-3.0}$ & $42\%$ \tabularnewline
$[280, 300]\,\mathrm{G}\lambda$ & $0.93^{+0.02}_{-0.02}$ & $17.3^{+0.8}_{-1.5}$ & $69\%$ \tabularnewline
\hline
\hline 
\end{tabular}
}\caption{Spin-inclination inference results for the survey over the $140$ emission profiles listed in Tab.~\ref{tbl:Parameters} with underlying black hole spin $a_{*}=0.94$ and inclination $i_{*}=17^{\circ}$. We use a Gaussian kernel density estimator on the marginalized spin and inclination distributions shown in Fig.~\ref{fig:SpinInclinationDistribution} to estimate a $2\sigma$ confidence interval for each, listed as $\pm$ values around the central $50\%$ parameter estimate. The errors quoted here are sourced by varying the model of emission (and not due to measurement uncertainties as considered in Sec.~\ref{sec:NoiseStudy}). We also list the percentage of emission profiles for which we successfully inferred a spin-inclination pair.}\label{tbl:AstroSpinIncInferenceStatistics}
\end{table}

We find that at short baselines ($u\lesssim100\,\mathrm{G}\lambda$), the version of the method which infers both spin and inclination performs markedly worse than that which infers spin alone (e.g., the results presented in Sec.~\ref{sec:AstroStudy}). The reason for this appears to be that the inferred ring diameter $d_{||}$ at these baselines is far less robust against changes in the emission profile than the fractional shape asymmetry. 

The confidence ellipses shown in the first two panels in Fig.~\ref{fig:SpinInclinationDistribution} are angled diagonally and the distribution of the inferred spin-inclination pairs appear to lie on a curve (or a collection of curves forming a tight band) whose shape looks similar to the fixed asymmetry contours in Fig.~\ref{fig:DiameterAsymmetryContours}. This suggests that for each of the successful inferences across the baseline windows $[20, 40]\,\mathrm{G}\lambda$ and $[80, 100]\,\mathrm{G}\lambda$, respectively, the inferred fractional asymmetry is, in relative terms, clustered closely around some value across the different emission profiles. Meanwhile, the larger variations in the inferred diameter cause the contour-like structure of each distribution which looks similar to fixing a (red) asymmetry contour level in Fig.~\ref{fig:DiameterAsymmetryContours} and plotting the points of intersection with each of the (blue) diameter contours therein.

For inferences across the baseline window $[280, 300]\,\mathrm{G}\lambda$, the spin-inclination inference scheme performs far better. Similar to spin-inference in Sec.~\ref{sec:AstroStudy}, the inferred spin confidence interval $a=0.93^{+0.02}_{-0.02}$ is centered almost exactly on the true spin value $a_{*}=0.94$, while here we also obtain an inferred inclination $i={17.3^{\circ}}^{+0.8^{\circ}}_{-1.5^{\circ}}$ which is also close to the true value $i_{*}=17^{\circ}$. Both parameter inference methods successfully provide sharp estimates for $69\%$ of the emission profiles considered.

Taken together, the results presented in this appendix and Sec.~\ref{sec:AstroStudy} suggest that the fractional asymmetry inferred at baselines where the $n=1$ ring typically dominates (for the emission profiles here considered) is much more robust than the diameter $d_{||}$ against changes in the emission model, and thus more closely tracks the critical curve asymmetry than the inferred $d_{||}$ does the parallel diameter of the critical curve.

\newpage
\bibliographystyle{apsrev4-1}
\bibliography{refs}
\end{document}